\documentstyle[twoside,11pt]{article}

\newtheorem{theorem}{Theorem}[section]

\newtheorem{lemma}[theorem]{Lemma}
\newtheorem{conjecture}[theorem]{Conjecture}

\begin{document}
\title{Dynamical Systems: Some Computational Problems}
\author{John Guckenheimer\thanks{Research partially supported by
	the National Science Foundation.}
 and Patrick Worfolk\footnotemark[1] \\
        Center for Applied Mathematics \\
        504 Engineering and Theory Center \\
        Cornell University \\
	Ithaca, NY 14853 \\
	USA}
\date{}
\maketitle

\begin{abstract}
We present several topics involving the computation of dynamical
systems.  The emphasis is on work in progress and the presentation is
informal -- there are many technical details which are not fully
discussed.  The topics are chosen to demonstrate the various
interactions between numerical computation and mathematical theory in
the area of dynamical systems.  We present an algorithm for the
computation of stable manifolds of equilibrium points, describe the
computation of Hopf bifurcations for equilibria in parametrized
families of vector fields, survey the results of studies of
codimension two global bifurcations, discuss a numerical analysis of
the Hodgkin and Huxley equations, and describe some of the effects of
symmetry on local bifurcation.
\end{abstract}

\section{Introduction}

This article is a written record of the lectures delivered at the 1992
NATO Summer School in Montreal. The spirit of the lectures was
informal, and we have tried to preserve the informality and didactic
quality of the lectures. The lectures dealt with several related
topics, all involving computations of dynamical systems. The emphasis
here stresses topics that have not yet been fully developed in other
recent papers.  Accordingly, they reflect to a large extent ``work in
progress'' rather than presenting a summary and review that makes a
finished mathematical tale. There is also a lack of rigor that
reflects a pragmatic approach that seeks reliable answers to questions
even when these answers cannot be totally integrated into formal
mathematical theory.

Our goal is to answer questions about specific dynamical systems. If
$f : R^n \times R^k \rightarrow R^n$ defines a $k$ parameter family of
vector fields, we would like to know everything about the phase
portraits of $f$ and the bifurcations that occur as parameters are
varied.  Since most dynamical systems cannot be integrated in terms of
analytic expressions that are valid for all time, this is a task that
can only be approached sensibly through numerical computation. This is
problematic for mathematicians because naive error estimates tend to
grow exponentially with iterative calculations such as the numerical
integration of a vector field. For the most part, we ignore this
difficulty and proceed with confidence that computers provide good
approximations to trajectories of vector fields. We seek to find
additional algorithms that allow us to calculate aspects of dynamical
systems that are hard to determine otherwise. Though these lectures do
not emphasize applications, our interests extend beyond individual
algorithms to the implementation of efficient and powerful problem
solving environments for the analysis of dynamical systems and their
bifurcations \cite{dstool:AMS}.

\section{Computing Stable Manifolds of Equilibria}

The stable and unstable manifolds of equilibrium points are important
geometric objects in the phase portraits of vector fields. The Stable
Manifold Theorem asserts their existence, and some proofs of the
Stable Manifold Theorem are constructive; e.g. \cite{PdM}.
Nonetheless, the numerical computation of stable and unstable
manifolds of dimension larger than one presents difficulties. This
section describes some of these difficulties and sketches an approach
that leads to a reasonable algorithm for displaying the
two-dimensional stable manifold of the Lorenz system
\cite{Lorenz}.

We begin by recalling the Stable Manifold Theorem for equilibria of
finite-dimensional vector fields:

\begin{theorem}
Let $X$ be a $C^r$ vector field on an n-dimensional manifold $M$ with
an equilibrium point $p$ and flow $\Phi$.  Let $L:T_p(M) \rightarrow
T_p(M)$ be the linearization of $X$ at $p$.  Assume that $T_p(M)$
splits as an invariant direct sum $ E^s + E^u $ with the spectrum of
$L|E^s$ in the open left half plane of $\cal{C}$ and the spectrum of
$L|E^u$ in the open right half plane of $\cal{C}$. Then there are
$C^r$ submanifolds $W^s$ and $W^u$ passing through $p$ with tangent
spaces $E^s$ and $E^u$ respectively with the following property: $$W^s
= \{x \in M | \Phi(t,x) \rightarrow p \mbox{ as } t \rightarrow
\infty\}$$ $$W^u = \{x \in M | \Phi(t,x) \rightarrow p \mbox{ as } t
\rightarrow - \infty\}\;.$$ If $\Phi$ is complete, then $W^s$ and
$W^u$ are $1-1$ immersions of Euclidean spaces into $M$.
\end{theorem}

There is a naive approach to computing the stable manifold of an
equilibrium that can be derived from the simplest proofs of the Stable
Manifold Theorem. If $p$ is an equilibrium point with the
linearization at $p$ having stable manifold $E^s$ and unstable
manifold $E^u$, the stable manifold of $p$ can be represented in local
coordinates as the graph of a function $\gamma:E^s \rightarrow E^u$.
Near $p$, each affine subspace parallel to $E^u$ intersects $W^s$ in a
single point. The function $\gamma$ has vanishing derivative at $p$,
so points of $E^s$ near $p$ form a good $C^1$ approximation to $W^s$
there.  Using the invariance of the stable manifold, the backwards
trajectories starting on a sphere surrounding $p$ in $E^s$ sweep out
an approximation to $W^s$. One can choose a set of initial conditions
on $E^s$ and compute their trajectories to visualize $W^s$, but the
process often works poorly.

Let us examine the Lorenz system as an example. The equations of the
Lorenz system are
\begin{eqnarray*}
\dot{x} & = & \sigma(y-x) \\
\dot{y} & = & \rho x - y - x z \\
\dot{z} & = & -\beta z + xy  \\
\end{eqnarray*}
with ``standard'' parameters $\sigma = 10$, $\beta = 8/3$, and $\rho =
28$.  For these parameters, the origin is a saddle point with a
two-dimensional stable manifold and a one-dimensional unstable
manifold. The linearization at the origin is given by the matrix
\[ \left( \begin{array}{ccc}
-10&10&0 \\
28&-1&0 \\
0&0&-8/3
\end{array} \right) \;. \]
The negative eigenvalues of this matrix are approximately $-2.67$ and
$-22.8$ with a ratio that is approximately $8.56$. There are two
problems associated with the large ratio between these eigenvalues.
First, ``most'' trajectories in the stable manifold approach the
origin from directions close to the $z$ axis which is the
eigendirection of the eigenvalue $-8/3$. This can be seen readily by
solving the linear system
\begin{eqnarray*}
\dot{x} & = & - a x \\
\dot{y} & = & - b y \;\;\;.
\end{eqnarray*}
The trajectories of this system lie along the curves $y = c x^{a/b}$.
Points uniformly distributed on the unit circle approach the origin
with much higher density close to the $y$ axis if $a \ll b$. The
second problem is that the trajectories flow much faster in the
eigendirection of the eigenvalue of larger magnitude than in the
eigendirection of the weaker eigenvalue. A small circle of initial
conditions in the stable manifold will flow backwards to an ellipse
whose axes have a ratio approximately $e^{8.56} \approx 5220$ after
one unit of time. Thus a fixed time integration of the vector field on
a set of initial conditions clustered near the weak eigendirection
still will yield a set of points that stretches along the strong
eigendirection.

This discussion indicates that strategies for drawing stable manifolds
based solely upon the numerical integration of trajectories starting
from initial values near the equilibrium may be ineffective.  To
obtain approximations to large regions of a stable manifold, our
approach is to mimic the construction of ``geodesic coordinates'' in
differential geometry. We describe the general construction.

Let $p$ be a hyperbolic equilibrium point of a vector field $X$ on a
Riemannian manifold $M$ and let $W = W^s(p)$ be the stable manifold of
$p$. We make use of the Riemannian metric induced on $W$ from its
embedding in the ambient Riemannian manifold M.  There is a
neighborhood $U$ of $p$ in $W$ such that $U-\{p\}$ is foliated by
spheres $S_r$ which lie at a distance $r$ from $p$ in the Riemannian
metric.  We would like to write a differential equation for the
evolution of these spheres with increasing $r$, and this is easy to do
when the vector field $X$ is not tangent to $S_r$.  The fundamental
observation that allows us to proceed is that the tangent space
$T_x(W)$ to $W$ at $x\in S_r$ is spanned by $X$ and $T_x(S_r)$ as long
as the vector field is not tangent to $S_r$.  The unique geodesic in
$W$ from $p$ to any point of the sphere $S_r$ is characterized by the
requirement that it is orthogonal to $S_r$ and that its tangent has
unit length and lies in the subspace spanned by the tangents to $S_r$
and the vector field $X$.  This provides the motivation for an
algorithm to evolve the sphere $S_r$ by extending the geodesics
through $p$ and points on the sphere, as pictured in Figure
\ref{geodesic}.
\refstepcounter{figure} \label{geodesic}
\begin{theorem}
Let $X$ be a $C^r$ vector field on an n-dimensional Riemannian
manifold $M$ with a hyperbolic equilibrium point $p$ and flow $\Phi$.
Let $W$ be the $s$-dimensional stable manifold of $p$ and let $S_r$ be
the geodesic sphere at distance $r$ from $p$ in the Riemannian metric
induced on $W$.  If $X$ is not tangent to $S_r$, then the vector field
obtained by orthogonally projecting $X$ onto the normal bundle of
$S_r$ is tangent to the geodesic rays of $W$ emanating from $p$.
\end{theorem}

There may well be tangencies between the vector field $X$ and the
geodesic spheres $S_r$ in a stable manifold. When this happens, it is
no longer a simple matter to ``evolve'' $S_r$ as a submanifold without
additional geometric information.  We approach the problem with the
question of determining when does the set of trajectories passing
through a submanifold $S$ of initial conditions form a smooth
submanifold $W$.  We point out two difficulties.  The following
example illustrates one problem that occurs if there are isolated
points of tangency between $X$ and $S$.

Let $X$ be the vector field $X = \partial_x - x \partial_y + x^2
\partial_z$ in $R^3$ and let $S$ be the $x$ axis. The vector field $X$
is tangent to $S$ at the origin. The flow of $X$ is defined by the map
$\Phi(x,y,z,t) = (x+t,y-x t-t^2/2,z+x^2 t+x t^2 + t^3/3)$. The surface
of trajectories with initial conditions on the curve $S$ is given by
the map $F(x,t) = (x+t,-x t - t^2/2, x^2 t + x t^2 + t^3/3)$. The
image of $F$ is a singular surface whose intersection with the plane
$x=0$ is a cusp.  This example indicates that given a smooth
submanifold $S$ of dimension $s-1$, there are simple vector fields
with the property that the trajectories passing through the
submanifold do not form a smooth surface of dimension $s$.  Locally, a
submanifold $S$ may be evolved by a vector field $X$ in such a
fashion, as long as the tangent spaces to $S$ do not contain $X$.
However, at points of tangency additional information is required to
ensure that the trajectories through $S$ will form a smooth
submanifold.

There is a further difficulty that we encounter in evolving
submanifolds.  Suppose that $W$ is a submanifold with boundary and
that $X$ is a vector field with a segment of a trajectory $\gamma$ in
the boundary of $W$.  Assume that $W$ is contained in the interior of
an invariant manifold $\hat{W}$ of the same dimension.  It is not true
that $W$ and the vector field define $\hat{W}$.  If the trajectory
$\gamma$ never enters the interior of $W$, then $\hat{W} - W$ may
consist of points whose trajectories do not intersect $W$ at all. For
example, the most dramatic case occurs when the dimension of $W$ is
two and the boundary of $W$ is a periodic orbit.  In this case, we
cannot ``grow'' $W$ in the fashion that we wish to use for stable and
unstable manifolds.  On the other hand, the scenario we have just
described cannot occur for a stable or unstable manifold, since these
invariant manifolds are formed by a set of trajectories that are
asymptotic to an equilibrium point.

This story ends in an unfinished state. We would like to use
algorithms based upon computing geodesics for computing stable and
unstable manifolds because they display the geometry of the manifolds
in a way that naive computation of trajectories does not. On the other
hand, there are mathematical difficulties with the formulation of
algorithms that are based solely upon the computation of a vector
field that will be tangent to geodesics. This is an area of research
in which more work is warranted. The goal is to find a procedure that
``works'' for the largest possible class of examples, where the
criterion of success is the ability to carry out the computation to a
specified precision in a reasonable amount of time.
Figure~\ref{twodman} shows a picture of the computed stable manifold
for the origin of the Lorenz system.
\refstepcounter{figure} \label{twodman}

\section{Computing Hopf Bifurcations}

This section describes joint work in progress with Mark Myers and
Bernd Sturmfels. The question we discuss involves the computation of
points where Hopf bifurcations occur in parametrized families of
vector fields. Let $\dot{x} = f_\lambda(x)$ be a system of
differential equations on $R^n$ depending upon a $k$-dimensional
parameter $\lambda$. One of the first steps in a bifurcation analysis
of this family is the determination of parameter values at which there
are non-hyperbolic equilibrium points $(x,\lambda) \in R^n \times
R^k$.  These points are characterized by the conditions that
$f_\lambda(x) = 0$ and $Df_\lambda(x)$ has a zero or purely imaginary
eigenvalue.  In both cases, the set of parameter values at which
bifurcation occurs is expected to have codimension one in the
parameter space.  Generically, the case of a zero eigenvalue is a
saddle-node bifurcation while the case of pure imaginary eigenvalues
is a Hopf bifurcation \cite{GH}. Note that solving the equations
describing either type of bifurcation does not involve numerical
integration of the differential equations. Instead, one has a
root-finding problem to solve.  Since the Jacobian $Df_\lambda(x)$
depends upon $f$, it is not immediately apparent that one can express
this problem as a non-degenerate system of $n+1$ equations in $n+k$
variables.

For saddle-node bifurcations, zero eigenvalues of the Jacobian can be
detected by computing the determinant of the Jacobian. It is not
difficult to prove that the map $F:R^n \times R^k \rightarrow R^{n+1}$
given by $F(x, \lambda) = (f_\lambda(x), det(Df_\lambda(x))$ is
non-degenerate for generic $f$ at most points \cite{Keller}. The
fundamental example is given by the normal form for a saddle-node
bifurcation: $f_\lambda(x) = \lambda + x^2$.  In this example, $F(x,
\lambda) = (\lambda + x^2, 2x)$ is non-singular. The general
saddle-node bifurcation can be transformed to this normal form by
using the techniques of singularity theory \cite{GG}. In implementing
algorithms for computing saddle-node bifurcations, we face questions
about the efficiency and accuracy with which the determinant of the
Jacobian of a map can be computed.

We seek algorithms that will locate Hopf bifurcations in a manner
analogous to the computation of saddle-node bifurcations. Thus, we
want a map $H:R^n \times R^k \rightarrow R^{n+1}$ which vanishes when
$Df$ has an equilibrium at which there are a pair of pure imaginary
eigenvalues. A basic aspect of this problem is finding a criterion for
determining when a matrix has a pair of pure imaginary eigenvalues.  A
complete algebraic solution to this problem is more complex than
finding a criterion for zero eigenvalues. The set of matrices with
pure imaginary eigenvalues does not define an algebraic hypersurface
in the space of matrices, but rather a semialgebraic set. Let us see
what this means in concrete terms for $2 \times 2$ matrices. A $2
\times 2$ matrix has pure imaginary eigenvalues if and only if its
trace vanishes and its determinant is positive. Thus the set of $2
\times 2$ matrices with pure imaginary eigenvalues is specified by one
equation {\em and} one inequality on the coefficients of the matrix.
If we drop the inequality, then we find matrices with real eigenvalues
of equal magnitude and opposite sign as well as matrices with pure
imaginary eigenvalues.

There are classical algebraic theories that address the question of
when a matrix has a pair of pure imaginary eigenvalues. These theories
are now best known in the form of the Routh-Hurwitz criteria that a
matrix have all of its eigenvalues in the left half plane. We have
applied these theories and continuation strategies in order to
implement algorithms for computing Hopf bifurcations. Here we give a
brief indication of the algebraic foundation for our algorithms that
give criteria for a square matrix to have a pair of pure imaginary
eigenvalues. For the remainder of this section, we let $A$ denote an
$n \times n$ real matrix and $P(\lambda) = det(\lambda I - A)$ denote
its characteristic polynomial. The roots of $P$ are the eigenvalues of
$A$. There are two approaches to our algebraic problem: To give
criteria for the polynomial $P$ to have a pair of pure imaginary
roots, or, alternatively, to give criteria for $A$ to have a pair of
pure imaginary eigenvalues by performing algebraic transformations
directly on $A$.

We make the basic observation that if a real polynomial $P$ has a pure
imaginary root $\lambda$, then $-\lambda$ is also a root. Thus a
necessary condition for $P$ to have a pair of pure imaginary roots is
that the polynomials $P$ and $Q(x) = P(-x)$ have a common root. The
{\em Sylvester resultant} is a function of the coefficients of $P$ and
$Q$ that vanishes if and only if $P$ and $Q$ share a common root.
Applied in the context of our problem, we first reduce the size of the
problem by observing that the polynomials $R = P+Q$ and $S = P-Q$ are
even and odd, respectively.  Moreover, $P$ and $Q$ have a common root
if and only if $R$ and $S$ have a common root.  We can write $R(x) =
\hat{R}(x^2)$ and $S(x) = x\hat{S}(x^2)$ with $\hat{R}$ and $\hat{S}$
polynomials of degree approximately $n/2$.  Therefore, the Sylvester
resultant of $\hat{R}$ and $\hat{S}$ gives a function that vanishes if
and only if $P$ and $Q$ have a common root or, equivalently, $P$ has
two roots whose sum is zero. If we denote
\[
P(\lambda) = c_{_0} + c_{_1} \lambda + \cdots + c_{_{n-1}} \lambda^{^{n-1}}
+ \lambda^{^{n}} \; \; ,
\]
and $n$ is even,
the Sylvester resultant is the determinant of the matrix
\[
{\cal S} =
\mbox{
$\left ( \begin{array}{cccccccccc}
c_0 & c_2 & \cdots & c_{_{n-2}} & 1 & 0 & 0 & \cdots & 0 \\
0   & c_0 & c_2 & \cdots & c_{_{n-2}} & 1 & 0 & \cdots & 0 \\
\mbox{\normalsize $\vdots$} & & &  & & &  &  & \mbox{\normalsize $\vdots$} \\
0 & \cdots &\cdots &  0 & c_0 & c_2 & \cdots  & c_{_{n-2}} & 1 \\
c_1 & c_3 & \cdots & c_{_{n-1}} & 0 & 0 & \cdots & \cdots & 0 \\
0   & c_1 & c_3 & \cdots & c_{_{n-1}} & 0 & \cdots & \cdots & 0 \\
\mbox{\normalsize $\vdots$} &  &  & & & & & &  \mbox{\normalsize $\vdots$} \\
0 & \cdots  & \cdots & 0 & c_1 & c_3 & \cdots & \cdots & c_{_{n-1}}
\end{array} \right )
\begin{array}{ll}
      \left . \rule{0in}{.42in} \right \} & \frac{n-2}{2} \; \mbox{rows} \\
      \left . \rule{0in}{.42in} \right \} & \frac{n}{2} \; \mbox{rows}
\end{array}
$}
\]
while if $n$ is odd, it is the determinant of the  matrix
\[
\rule{.3in}{0.in} {\cal S} =
\mbox{
$\left ( \begin{array}{cccccccccc}
c_0 & c_2 & \cdots & c_{_{n-3}} & c_{_{n-1}} & 0 & 0 & \cdots & 0 \\
0   & c_0 & c_2 & \cdots & c_{_{n-3}} & c_{_{n-1}} & 0 & \cdots & 0 \\
\mbox{\normalsize $\vdots$} & & &  & & &  &  & \mbox{\normalsize $\vdots$} \\
0 & \cdots &\cdots &  0 & c_0 & c_2 & \cdots  & c_{_{n-3}} & c_{_{n-1}} \\
c_1 & c_3 & \cdots & c_{_{n-2}} & 1 & 0 & \cdots & \cdots & 0 \\
0   & c_1 & c_3 & \cdots & c_{_{n-2}} & 1 & 0  & \cdots & 0 \\
\mbox{\normalsize $\vdots$} &  &  & & & & & &  \mbox{\normalsize $\vdots$} \\
0 & \cdots  & \cdots & 0 & c_1 & c_3 & \cdots & c_{_{n-1}} & 1
\end{array} \right )
\begin{array}{ll}
      \left . \rule{0in}{.42in} \right \} & \frac{n-1}{2} \; \mbox{rows} \\
      \left . \rule{0in}{.42in} \right \} & \frac{n-1}{2} \; \mbox{rows}
\end{array}
\; \; . $}
\]

The resultant encodes the outcome of applying the Euclidean algorithm
to a pair of polynomials. If two polynomials have a common root, then
the Euclidean algorithm yields the greatest common denominator of the
two polynomials. The coefficients of this GCD can be expressed as
determinants in the coefficients of the two polynomials. These
determinants are the subresultants of the two polynomials.  These
resultants give explicit functions that define the locus of matrices
that have eigenvalues whose sum is zero. They do not determine whether
the roots are real or complex. For this purpose, one can use the
theory of {\em subresultants} \cite{Loos}.  We observe that a common
root of $P$ and $Q$ is imaginary if and only if the corresponding
common root of $\hat{R}$ and $\hat{S}$ is negative. Therefore, we can
give explicit inequalities in terms of the subresultants that
determine whether a polynomial with a single, simple pair of roots
whose sum is zero has a pair of pure imaginary roots. Table
\ref{hopftable} gives a list of the equations and inequalities that
determine the polynomials $P$ of degree at most six with these
properties.
\begin{table}
\centering
\begin{tabular}{|r|c|} \hline
\rule[-.08in]{0in}{.25in}
$n$ & $P(\lambda)=c_0+c_1\lambda+\ldots + c_{n-1}\lambda^{n-1}
	+ \lambda^n$ \\
\hline \hline
\rule[-.08in]{0in}{.25in}
$2$ & $c_1=0 \;,\;\; c_0>0$ \\
\hline
\rule[-.08in]{0in}{.25in}
$3$ & $c_0 - c_1 c_2 =0 \;,\;\; c_1>0$   \\
\hline
\rule[-.08in]{0in}{.25in}
$4$ & $c_0 c_3^2 - c_1 c_2 c_3 + c_1^2=0 \;,\;\; c_1 c_3>0$ \\
\hline
\rule[-.20in]{0in}{.45in}
$5$  & $\begin{array}{c} (c_2-c_3 c_4)(c_1 c_2 - c_0 c_3) + c_1 c_4(c_1 c_4 -
2 c_0) + c_0^2=0 \; , \\ (c_2 - c_3 c_4) (c_0 - c_1 c_4)>0 \end{array}$ \\
\hline
\rule[-.30in]{0in}{.65in} $6$ & $\begin{array}{c}
			 c_0 c_5^2 ( c_0 c_5 - c_2 c_3 ) +
			 c_1 c_5^2 ( c_2^2 - c_0 c_4 )+
			c_1 ( c_1^2 + c_0 c_3 c_5 ) +
			 c_1 c_5 ( c_0 c_3 - 2 c_1 c_2 ) + \\
			 ( c_4 c_5 - c_3 ) ( c_0 c_3^2  -
					c_0 c_1 c_5 +
				c_1^2 c_4 - c_1 c_2 c_3 ) =0 \; ,  \\
			  ( c_1 c_3 + c_0 c_5^2 - c_1 c_4 c_5 )
			  ( c_3^2 - c_1 c_5 + c_2 c_5^2 - c_3 c_4 c_5 ) >0
				\end{array}$ \\
\hline
\end{tabular}
\caption{ \label{hopftable} Conditions for $P(\lambda)$ to have a
pair of pure imaginary roots.}
\end{table}

The computation of characteristic polynomials of square matrices is
computationally expensive and prone to numerical errors for some types
of matrices. Thus, we seek methods that allow us to determine whether
a matrix has a pair of pure imaginary eigenvalues without computing
the characteristic polynomial as an intermediate step. This can be
done through the definition of appropriate Kronecker or tensor
products. The basic algebraic idea is that there are transformations
of matrices that produce matrices whose eigenvalues are functions of
the eigenvalues of the original matrix. For example, if $A$ and $B$
are square matrices, then the eigenvalues of the tensor product $A
\otimes B$ are the pairwise products of the eigenvalues of $A$ and
those of $B$. If we form the transformation $T = I \otimes A + A
\otimes I$ with $I$ the $n \times n$ identity matrix, then the
eigenvalues of $T$ are sums of pairs of eigenvalues of $A$. To remove
the redundancy associated with having both pairs $\lambda_i +
\lambda_j$ and $\lambda_j + \lambda_i$ as eigenvalues, we use a
skew-symmetric version construction. The {\em bialternate product} $A
\odot I$ of $A$ is the $n(n-1)/2 \times n(n-1)/2$ matrix defined by
\begin{eqnarray*}
\label{eq25}
\left ( 2 {\bf A}\! \odot {\bf I}_n \right )_{\{pq,rs\}} =
\left\{ \begin{array}{cl}
 -(A)_{p,s} & \mbox{if}  \; r=q \\
  (A)_{p,r} & \mbox{if}  \; r\not= p \; \mbox{and} \; s=q \\
  (A)_{p,p} + (A)_{q,q} & \mbox{if}  \; r=p \; \mbox{and} \; s=q \\
  (A)_{q,s} & \mbox{if}  \; r=p \; \mbox{and} \; s \not= q  \\
 -(A)_{q,r} & \mbox{if}  \; s=p \\
   0       & \mbox{otherwise}
\end{array} \right .
\end{eqnarray*}
where the rows are labeled {\em lexicographically} by $pq$ for
$(p=2,\ldots,n;q=1,\ldots,p-1)$ and the columns likewise by $rs$ for
$(r=2,\ldots,n;s=1,\ldots,r-1)$ \cite{Fuller}.  The eigenvalues of $A
\odot I$ are pairwise sums $\lambda_i + \lambda_j$ with $i<j$ of the
eigenvalues of $A$. Thus, a necessary condition for $A$ to have a pair
of pure imaginary eigenvalues is that $A \odot I$ be singular. This
singularity can be tested by computing the determinant of $A \odot I$,
but we can also use other algorithms (such as singular value
decomposition) from numerical linear algebra that give more accurate
and robust tests for the singularity of $A$. In our ongoing work, we
are implementing algorithms for tracking Hopf bifurcations based upon
these algebraic methods and standard continuation methods.

\section{Homoclinic Bifurcations and Their Computation}

Bifurcation theory describes qualitative changes in phase portraits
that occur as parameters are varied in the definition of a dynamical
system. For dynamical systems defined by vector fields on $R^n$, one
has a system of equations of the form $\dot{x} = f_\lambda(x)$ with
$\lambda \in R^k$ denoting a $k$-dimensional vector of parameters.  A
rough classification of bifurcations distinguishes between {\em local}
and {\em global} bifurcations, but it is difficult to make this
distinction precise. Heuristically, the dynamics of local bifurcations
are determined by information contained in the germ of the Taylor
series of $f$ or a Poincar\'{e} return map at a point, whereas the
dynamics of global bifurcations require information about the vector
field along an entire (non-periodic) trajectory.  While the theory of
bifurcations in multi-parameter families is far from complete, the
theory of global bifurcations is more fragmentary than that of local
bifurcations. A number of ``codimension 2'' global bifurcations have
been studied, but there has not been an attempt to construct a
synthesis of these studies that portrays a systematic view of the
different cases and phenomena that occur. The lectures that were given
in Montreal at this NATO sponsored summer school touched upon these
matters, but these notes go much farther towards the construction of
such a synthesis. We focus our attention on global bifurcations that
involve homoclinic or heteroclinic orbits of equilibrium points of a
vector field. We shall call homoclinic and heteroclinic orbits {\em
connecting} orbits in order to have a common name for the two.

\u{S}ilnikov seems to be the principal originator of the strategy we adopt
to study global bifurcations with connecting orbits in higher
dimensional vector fields. For planar vector fields, the techniques
are older and asymptotic methods for studying global bifurcations in
planar vector fields were well developed by the 1920's
(Dulac~\cite{Dulac}).  The fundamental idea is that the recurrent
behavior near a connecting orbit should be studied in a fashion
similar to that used in studying periodic orbits via a Poincar\'{e}
return map. In particular, codimension one cross sections to the flow
are introduced, and the return map of these cross sections is studied.
There are some additional complications in the study of connecting
orbits compared to that of periodic orbits which significantly
complicate the analysis. We mention three of these. First, the
discrete maps that describe flow past an equilibrium point are
singular. In the simplest cases of flow past non-resonant equilibrium
points, these ``passage'' or ``correspondence'' maps have
singularities of the form $x^a$ where $a$ is not an integer and may be
complex if the equilibrium has complex eigenvalues. These
singularities lead to analysis that is much more complicated and
intricate than that associated with the return map of a periodic
orbit. As the work presented by Ilyashenko and \'{E}calle at this
summer school demonstrated, this analysis is frightfully complicated
even for planar vector fields.  The second complication arising from
connecting orbits is that there are discontinuities in the return maps
of cross sections that are associated with connecting orbits. The
bifurcations and attractors that appear in the Lorenz system
\cite{Lorenz} give a vivid example of the consequences of these
discontinuities and have been described by Guckenheimer and Williams
\cite{GW}.  These discontinuities force one to look at dynamical
systems that are built from multiple pieces rather than studying the
iterates of a single continuous mapping. The third complication is
that the list of cases of qualitatively different type is
substantially longer than with local bifurcations. Tools for treating
all of the cases in a single analysis are lacking, so the construction
of a comprehensive and complete theory seems like a daunting task.

These lectures adopt a superficial view of the mathematical
technicalities associated with global bifurcations. In constructing
return maps for a flow along connecting orbits, one would like the
simplest possible analytical expressions for these return maps. Normal
form theory explores how coordinate changes can be used to simplify
these analytical expressions, but the theory produces a large and
intricate story whose details can mask many of the phenomena that we
want to study. The idea of \u{S}ilnikov was to relax the requirement
that an exact normal form be used to describe passage past an
equilibrium and to use an approximation to the normal form instead.
Asymptotic expansions for the flows and passage maps can be
constructed, and one can begin the analysis of bifurcations with
connecting orbits by using the leading term, or terms, of these
asymptotic expansions. A similar approach to the global part of the
flow producing returns to a cross section can also be used, but the
asymptotic expansions of these smooth maps are given simply by Taylor
series. A matching procedure can then be used to represent the full
return map as a composition of singular and regular maps that come
from passage past equilibrium and parts of the flow that are
non-singular. In constructing maps with such matching procedures, we
must remember that there may be discontinuities that lead to the
examination of several different sequences of compositions. Following
an analysis of return maps built from approximations, we can seek to
determine the structural stability of the systems. It is unreasonable
to expect that each type of bifurcation will have a structurally
stable unfolding. We are plunged into the morass of considerations
that result from the fact that structurally stable vector fields are
not dense in the space of vector fields. The most that we try to do is
to explore in each case which dynamical features of the bifurcation do
persist under perturbations. Note that even specifying which
perturbations are to be allowed in a theory based upon the return maps
of connecting orbits is a tricky matter due to the singularities of
the maps. Rather than engaging the reader in an extensive discussion
of these problems, we proceed past them while erecting signposts that
point in the direction of unresolved and incomplete technical matters.

\subsection{Generic Homoclinic Orbits}

Let $F:R^{n} \rightarrow R^{n}$ be a sufficiently smooth, vector field
with a homoclinic orbit to an equilibrium point at $0$. We shall
denote the flow of $F$ by $\phi_t$ and the homoclinic orbit by $\gamma
(t)$.  Let
\[
\Gamma = \{ \gamma(t):t\in R\} \mbox{ where }
 \lim_{t \rightarrow \pm\infty} \gamma(t)=0 \mbox{ and } \gamma(t)\neq 0.
\]
We give a list of conditions that generic systems with a homoclinic
orbit satisfy. These conditions help ensure that bifurcations from the
homoclinic orbit have as simple a structure as possible. In
classifying codimension two bifurcations of homoclinic orbits, we can
then look at events that cause the failure of one of these genericity
conditions and examine the effect on bifurcations in two parameter
families.

The first condition is that the equilibrium be hyperbolic and
non-resonant.  The non-resonance conditions that are important are
primarily ones that involve the stable and unstable eigenvalues of
smallest magnitude. We call these eigenvalues the {\em principal
eigenvalues} of the equilibrium.  More precisely, assume the
linearization $D_xF(0)$ has $k$ eigenvalues in the left half plane and
$n-k$ in the right half plane.  Assume there exists a real principal
eigenvalue in the stable manifold denoted $\lambda_{s}$ or a complex
conjugate pair $\lambda_{s},\bar{\lambda}_{s}$ with the property that
if $\lambda$ is another stable eigenvalue of $D_xF(0)$, then
$\mbox{Re}(\lambda) < \mbox{Re}(\lambda_s) < 0$. Similarly, assume
there exists a real principal eigenvalue in the unstable manifold
denoted $\lambda_{u}$ or a complex conjugate pair
$\lambda_{u},\bar{\lambda}_{u}$ with the property that if $\lambda$ is
another unstable eigenvalue of $D_xF(0)$, then $0 <
\mbox{Re}(\lambda_u) < \mbox{Re}(\lambda)$.

At points in the stable and unstable manifolds of $0$, there are
filtrations of the tangent bundle associated with the exponential
growth or decay of vectors as the trajectory approaches the origin.
For a point $p$ of the stable manifold, we are interested in three
subspaces that we denote $E^{s+}(p) \supset E^s(p) \supset E^{ss}(p)$
and call the {\em stable plus weak unstable manifold}, the {\em stable
manifold} and the {\em strong stable manifold}, respectively.  These
are defined via the variational equation for the trajectory through
the point $p$ given by
\[ \dot{\xi} = (D_xF)|_{\phi_t(p)} \xi \; . \]
Let $V_t(v;p)$ denote the solution to this linear equation with
initial condition $V_0(v;p)=v \in T_pR^n$.
Let $\nu_s$ and $\nu_u$ be numbers such that
\[
\mbox{Re}(\lambda) < \nu_s < \mbox{Re}(\lambda_s) < 0 \; \mbox{ or } \;
 0 < \mbox{Re}(\lambda_u) < \nu_u < \mbox{Re}(\lambda).
\]
for any eigenvalue $\lambda$ of $D_xF(0)$ which is not a principal
eigenvalue. Then
\[
\begin{array}{c}
E^{s+}(p) \equiv \{ v \in T_pR^n | \lim_{t \rightarrow +\infty}e^{-\nu_u t}
|V_t(v;p)| = 0 \} \\
E^{s}(p) \equiv \{ v \in T_pR^n | \lim_{t \rightarrow +\infty}
|V_t(v;p)| = 0 \} \\
E^{ss}(p) \equiv \{ v \in T_pR^n | \lim_{t \rightarrow +\infty}e^{-\nu_s t}
|V_t(v;p)| = 0 \} \;.
\end{array}
\]
There are analogous definitions of the {\em unstable plus weak stable
manifold}, the {\em unstable manifold} and the {\em strong unstable
manifold} of a point p in the unstable manifold of $0$:
\[
\begin{array}{c}
E^{u+}(p) \equiv \{ v \in T_pR^n | \lim_{t \rightarrow -\infty}e^{-\nu_s t}
|V_t(v;p)| = 0\} \\
E^{u}(p) \equiv \{ v \in T_pR^n | \lim_{t \rightarrow -\infty}
|V_t(v;p)| = 0 \} \\
E^{uu}(p) \equiv \{ v \in T_pR^n | \lim_{t \rightarrow -\infty}e^{-\nu_u t}
|V_t(v;p)| = 0 \} \;.
\end{array}
\]
Note that there is no apparent relationship between these sets of
manifolds at a point $p$ of a homoclinic orbit other than that the
vector field $F(p)$ belongs to both $E^s(p)$ and $E^u(p)$.

Our next requirement for a generic homoclinic orbit involves the
direction of approach of a homoclinic orbit to $0$ as $t \rightarrow
\pm \infty$. Associated with the origin itself is a strong stable
manifold $W^{ss}$ consisting of points $p\in W^s$, the stable manifold
of the origin as defined by the Stable Manifold Theorem, for which the
vector field $F(p)$ lies in $E^{ss}(p)$. Note that
$T_0W^{ss}=E^{ss}(0)$ just as $T_0W^s=E^s(0)$.  Now $W^{ss}$ is a
proper submanifold of $W^{s}$ of codimension one or two depending upon
whether the principal stable eigenvalue is real or complex. In either
case, almost all trajectories in $W^s$ lie in $W^s \backslash W^{ss}$.
Similarly, almost all trajectories in $W^u$ lie in $W^u \backslash
W^{uu}$. Our next requirement for a generic homoclinic orbit is that a
point $p$ on $\Gamma$ satisfy
\[
p \not\in W^{ss},\; p \not\in W^{uu} \;.
\]
In words, (H3) is the statement that the homoclinic orbit approach the
origin in the directions of the principal eigenvectors.  This may
alternatively be expressed by requiring that a point $p$ on $\Gamma$
satisfy
\[
F(p) \not\in E^{ss},\; F(p) \not\in E^{uu} \; .
\]

The final condition that we impose upon generic homoclinic orbits
involves the derivative of the flow along the homoclinic orbit. We
state the condition in terms of the stable plus weak unstable and the
unstable plus weak stable manifolds.  Depending upon the types of the
principal eigenvalues, the sum of the dimensions of $E^{s+}(p)$ and
$E^{u+}(p)$ is $n+2$, $n+3$, or $n+4$. We require that the
intersection $E^{s+}(p) \cap E^{u+}(p)$ be transverse and,
furthermore, that $E^{s+}(p) \cap E^{uu}(p) = E^{ss}(p) \cap E^{u+}(p)
= \{0\}$.

Let us examine the meaning of this final condition in the case of real
eigenvalues.  Assume that $\lambda_s,\lambda_u\in R$.  Then
dim$(E^{ss})=m-1$ and dim$(E^{uu})=n-1$.  Along $\Gamma$, the
intersection $W^{s+} \cap W^{u+}$ is a two-dimensional bundle that
contains the vector field. At the origin, this bundle approaches the
plane spanned by the principal eigenvectors, both for $t \rightarrow
+\infty$ and for $t \rightarrow -\infty$. We can think of this bundle
as a ``ribbon'' along the homoclinic orbit that defines the behavior
of the system in the ``weak'' directions that determine the primary
structure of the orbit. Taking the closure at the origin of the bundle
along the homoclinic orbit, we obtain a bundle of planes along a
simple closed curve. This bundle is either orientable or
non-orientable. We distinguish these cases by calling them {\em
twisted} and {\em untwisted} homoclinic orbits. Twisted homoclinic
orbits cannot occur for vector fields on orientable two-dimensional
manifolds.  Both twist types of homoclinic orbits are represented
pictorially in Figure~\ref{homo}.
\refstepcounter{figure} \label{homo}

We will call a homoclinic orbit with real principal eigenvalues a {\em
binodal} homoclinic orbit.  For the generic binodal homoclinic orbit,
there are no additional interesting dynamical structures in a
sufficiently small neighborhood of the homoclinic orbit
\cite{Silnikov:generation}.  This is not necessarily true of generic
homoclinic orbits with a complex principal eigenvalue.  When exactly
one of the principal eigenvalues is complex, we will call the
homoclinic orbit a {\em unifocal} homoclinic orbit, and the generic
case has been studied by \u{S}ilnikov
\cite{Silnikov:countable,Silnikov:generation,Silnikov:contribution}
and others
\cite{Gaspard:generation,GaspardKapral:homoclinic,GS:h,Tresser:silnikov}.
The following two theorems describe the dynamical structures close to
a generic unifocal homoclinic orbit, assuming that $\lambda_u\in R$
and $\lambda_s\in C$.

\begin{theorem}
If $| \lambda_u / \mbox{Re}(\lambda_s)| > 1$ (\u{S}ilnikov condition)
then there exist horseshoes in every neighborhood of $\Gamma$.
\end{theorem}
\begin{theorem}
If $| \lambda_u / \mbox{Re}(\lambda_s)| < 1$ then
there is a neighborhood of $\Gamma$ which contains no
periodic orbits.
\end{theorem}

The {\em non-resonance condition} on the principal eigenvalues is
given by $| \lambda_u / \mbox{Re}(\lambda_s)| \neq 1$.  We may look at
further conditions on the eigenvalues and determine a second
genericity requirement: $| \lambda_u / \mbox{Re}(\lambda_s)| \neq 2$.
If $1 < | \lambda_u / \mbox{Re}(\lambda_s)| < 2$, then the linear flow
at the origin is contracting and the horseshoes are attracting, while
for $| \lambda_u / \mbox{Re}(\lambda_s)| > 2$ the linear flow at the
origin is expanding.

Finally we consider the generic homoclinic orbit with complex
principal eigenvalues which is called a {\em bifocal} homoclinic
orbit.  Figure \ref{focal} depicts the unifocal and bifocal homoclinic
orbits.  The bifocal homoclinic orbit has been studied by \u{S}ilnikov
\cite{Silnikov:existence,Silnikov:contribution}, Glendinning
\cite{Glendinning:subsidiary}, and Fowler and Sparrow
\cite{FowlerSparrow:bifocal}.  They prove the following theorem:
\begin{theorem}
There exist horseshoes in every neighborhood of a generic bifocal
homoclinic orbit.
\end{theorem}
\refstepcounter{figure} \label{focal}

\subsection{\u{S}ilnikov Coordinates and Local Normal Forms}

The unfoldings of the bifurcations involving connecting orbits can be
very complicated. As we stated above, certain types of generic
homoclinic orbits are embedded in much more complex dynamical
structures.  A full description of the unfolding of these orbits
entails a comprehensive analysis of the horseshoes that are created or
destroyed as a parameter is varied.  These unfoldings have been
studied for unifocal homoclinic orbits
\cite{Gaspard:generation,GaspardKapral:homoclinic,GS:h}
and bifocal ones
\cite{FowlerSparrow:bifocal,Glendinning:subsidiary}.
We do not undertake such a monumentally complicated task, but we do
want to describe maps that give an approximation to the return maps
associated with homoclinic orbits. The general procedure we employ for
doing this involves a decomposition of the return map for a homoclinic
orbit into two parts: one describing the ``local'' part of the flow
past the equilibrium point in the homoclinic orbit and the other
describing the ``global'' portion of the flow outside of this
neighborhood.  In cases of singular cycles containing more than one
connecting orbit, we make a decomposition into more pieces, but the
principle is the same. In the decompositions that we use, we try to
choose coordinates in a manner that simplifies the analytical
expressions of the vector fields.

The simplest flows near an equilibrium point are linear. The question
as to whether coordinate transformations near an equilibrium point can
be found that linearize a vector field in a neighborhood of the
equilibrium has been studied systematically for over a century,
starting with Poincar\'{e}'s dissertation.  The answers to the
question are complicated. See Arnold \cite{Arnold:geommethods} for an
extensive summary of what is known about the linearization problem.
Here we shall use a few bits of this theory.  {\em Resonance
conditions} on the eigenvalues provide the most elementary
obstructions to linearization. A resonance condition of order $k$ is
expressed in terms of eigenvalues $\lambda_j$ by $\lambda_i = \sum a_j
\lambda_j$ with the $a_j$ nonnegative integers whose sum is $k$. A
vector field with an equilibrium point that satisfies a resonance
condition of order $k$ usually cannot be linearized by a $C^k$
coordinate transformation.  Nonetheless, there are {\em resonant
normal forms} for the equilibrium with the property that there are
smooth coordinate changes that transform the degree $l$ Taylor series
of the vector field at the equilibrium to its resonant normal form.
We can hope in these situations that it is feasible to describe
explicitly the flow of the resonant normal forms truncated to degree
$l$ and that these flows serve as good approximations to the flow of
the original vector field in the vicinity of the equilibrium point.
Here the classical theory breaks down and does not provide a good
solution to the question as to when the passage map of two flows near
an equilibrium are good approximations to one another.

Consider a {\em linear} vector field $\dot{x}=f(x)$ with a hyperbolic
equilibrium point at the origin with $k$ stable eigenvalues and $n-k$
unstable eigenvalues.  Choose a coordinate system so that the stable
and unstable manifolds $W^s$ and $W^u$ of the origin lie in coordinate
planes.  Let $U_0$ be a neighborhood of the origin that is the product
of balls $B^s_r$ and $B^u_r$ of radius $r$ in the stable and unstable
manifolds $W^s$ and $W^u$.  The boundary of $B^s_r \times B^u_r $ is
$\partial B^s_r \times B^u_r \cup B^s_r \times \partial B^u_r $.
Since the vector field $f$ is linear, the motion of points is the
superposition of motions along the stable and unstable manifolds.
Furthermore, the distance to the origin of points in the stable
manifold decreases monotonically, while the distance to the origin of
points in the unstable manifold increases monotonically, in a well
chosen coordinate system.  Therefore the passage map of $U_0$ will be
defined as a map $\phi :\partial B^s_r \times B^u_r \rightarrow B^s_r
\times \partial B^u_r$.  With this choice of neighborhood of the
origin, it is difficult to determine an explicit expression for the
time of flight for a trajectory to reach the outgoing boundary of $U_0$
and hence to obtain a formula for $\phi$. On the other hand, if we
choose different neighborhoods of the origin, then we can
sometimes compute the exit time from the neighborhood. In the case of
real eigenvalues, we find that there are power law singularities;
complex eigenvalues produce singularities with other elementary functions.
Explicit examples are computed later.

For some problems of higher codimension, we shall need to compute the
passage maps of equilibrium points that are either resonant or
nonhyperbolic. In these cases, we shall still seek coordinate systems
and approximations for which there are explicit integrals for the
local normal forms at the equilibrium point. Cases with this property
are the only ones considered in this paper, though there are higher
codimension problems for which the normal forms are not integrable. In
finding passage maps for these resonant cases, we would like to obtain
formulas that remain valid when we perturb the equilibrium to make it
generic. This requires some additional care beyond merely solving the
explicit flows for the passage map at the resonant equilibrium.

The \u{S}ilnikov procedure is to combine the local passage maps near
equilibria with nonsingular maps that describe the flow between cross
sections around the portions of connecting orbits that do not contain
equilibria. In the case of a homoclinic orbit, these cross sections
can be taken to be portions of the boundary of the neighborhood $U_0$
that was used to construct the passage map past the equilibrium. As
with the passage maps, we seek approximations for these ``global''
maps between cross sections.  Since the maps are nonsingular, they can
be approximated by truncating their Taylor series.  Generally, we
start with affine approximations to the transformations.  If these are
inadequate to obtain the (structural) stability results we seek, then
higher degree approximations are used.  The return map for a cross
section can be obtained by composing these global maps with the local
passage maps. In carrying through this composition, care must be taken
with understanding the domains on which the maps are defined. When
there are multiple connecting orbits, there is the additional
possibility that flow in different directions away from an equilibrium
may produce structures that require following different patterns of
return to a neighborhood of the equilibrium.

\subsection{Bifurcations From a Homoclinic Orbit: What Can We Study?}

The difficulties of proving rigorous theorems about the unfoldings of
bifurcations are formid\-able. Nonetheless, we would like to prove as
much about phenomena that are consequences of bifurcations of
homoclinic and heteroclinic orbits as we can. Indeed, we would like to
capture the major dynamical events that occur in the unfolding despite
the fact that structural stability of the unfoldings will often fail.
To make the theory ``local'' to the bifurcating orbits, we restrict
attention to an arbitrarily small neighborhood of these orbits.
Recurrent behavior that involves trajectories that leave such a
neighborhood will be discussed separately.  Here we describe three
types of structures that can occur in small neighborhoods of
bifurcating connecting orbits.

The first type of dynamical structure that can be involved at a
homoclinic bifurcation is a periodic orbit.  Generic bifurcations of
binodal homoclinic orbits produce periodic orbits whose stability is
determined by the sum of the principal eigenvalues at the saddle. In
higher codimensions, multiple periodic orbits can bifurcate.  In some
cases (notably, the ``gluing bifurcation''
\cite{GGT}), these orbits and their patterns of bifurcations can be
complicated. The second type of dynamical structure produced in
bifurcations of connecting orbits is typified by the homoclinic
bifurcation in the Lorenz system \cite{Lorenz}.  In this system, for
reasons of symmetry, a pair of homoclinic orbits are formed. As they
bifurcate, a horseshoe is immediately created. The same phenomenon
occurs in higher codimension bifurcations without symmetry.  Third,
generic homoclinic orbits with complex values may be adjacent to
horseshoes. As described by \u{S}ilnikov \cite{Silnikov:countable},
homoclinic orbits to equilibria satisfying appropriate conditions on
their eigenvalues occur only in the closure of horseshoes. As the
homoclinic orbit is unfolded, the horseshoes closest to the homoclinic
orbit can be destroyed.

The singularity theory for mappings between two spaces is much
``cleaner'' than the bifurcation theory of flows (or iterated
mappings). There are a number of reasons for these differences, one
being the complex dynamical structures that occur in unfoldings of
certain bifurcations.  If there are horseshoes that are created or
destroyed as part of a bifurcation, then the unfolding of these
bifurcations will involve all of the associated complications. These
have been studied most thoroughly in the context of the H\'{e}non
mapping
\cite{BenedicksCarleson,MV}.
There are an infinite number of bifurcations of periodic orbits that
are part of the creation of horseshoes. If there is a single unstable
eigenvalue and volume contraction in these flows, there are also
phenomena such as infinitely many (Newhouse) sinks and the occurrence
of nonhyperbolic strange attractors \cite{GH}. We do not want to
become enmeshed in these details. To a large extent they appear to be
subsidiary to the fact that horseshoes are created, and we expect to
add little to the general discussion of the processes associated with
the creation and destruction of the horseshoes. Our focus in dealing
with horseshoes will be to demonstrate that a return map in an
unfolding satisfies the conditions that guarantee the existence of
horseshoes.  We will not generally explore whether the horseshoe is
part of a larger invariant set or investigate the presence of
nonhyperbolic invariant sets in the unfoldings.

\subsection{Codimension Two Bifurcations of Connecting Orbits}

There are many types of codimension two bifurcations of connecting
orbits.  Failure of one of the conditions that characterize a generic
homoclinic orbit will lead to a degenerate bifurcation.  These can be
classified into four groups:
\begin{enumerate}
\item
Eigenvalue degeneracies
\item
Degenerate approach
\item
Degenerate twist
\item
Multiple connecting orbits
\end{enumerate}

Each of these types can be further subdivided. For example, the
eigenvalue degeneracies may be due to a single zero eigenvalue, a pair
of pure imaginary eigenvalues, equal magnitude of the real parts of
the principal eigenvalues, or equal magnitude of a real principal
eigenvalue with the sum of a pair of imaginary principal eigenvalues.
Furthermore, one has a different analysis depending on whether the
homoclinic orbit is nodal, unifocal or bifocal, and, in the nodal
case, the twist associated with a resonance.  In all of the cases that
have been studied, the introduction of \u{S}ilnikov coordinates and
the study of return maps built from these coordinates is a major
portion of the analysis of the bifurcation. Rigorous results that go
beyond the description of model systems tend to be very difficult to
formulate and prove. When there are horseshoes associated with these
bifurcations, even the phenomenology associated with the \u{S}ilnikov
approximations tends to be incomplete.

Below we discuss a bifurcation with an eigenvalue degeneracy, one with
a degenerate approach, and another with multiple connecting orbits,
describing each in terms of \u{S}ilnikov approximations.
Additionally, within each category, we give a survey of other work
known to us.  For the case of degenerate approach, we refer the reader
to the work of Terman \cite{Terman:transition}.

\subsection{Eigenvalue Degeneracies}

There are a number of different eigenvalue degeneracies which may
occur.  Non-hyperbolic equilibrium point degeneracies have been
studied by, in the nodal case, Deng \cite{Deng:nonhyperbolic},
Luk'yanov \cite{Lukyanov:bifurcations} and Schecter
\cite{Schecter:saddlenode}; in the case of unifocal homoclinic orbits
with a nonhyperbolic saddle, Belyakov \cite{Belyakov:bifurcation};
and, in the case of unifocal homoclinic orbits with a nonhyperbolic
focus, Argoul, Arneodo and Richetti
\cite{ArgoulArneodo:experimental,RichettiArgoul:intermittency}, Bosch
and Simo \cite{BoschSimo:attractors}, Gaspard and Wang
\cite{GaspardWang:homoclinic}, and Hirschberg and Knobloch
\cite{HirschbergKnobloch:silnikovhopf}.  Here we will look at vector
fields that have a homoclinic orbit with real principal eigenvalues of
equal magnitude.  The homoclinic orbit may be twisted or untwisted,
and the bifurcation is different, depending on the case.  This has
been studied by a number of different authors, including Glendinning
\cite{Glendinning}, Kokubu \cite{Kokubu:homoclinic}, and Chow, Deng,
and Fiedler \cite{ChowDeng:resonant}.

A normal form for a planar saddle with a 1:1 resonance is given by the
system
\begin{eqnarray*}
\dot{x} & = & -x(1+xy) \\
\dot{y} & = & y
\end{eqnarray*}
An unfolding of this system allows the ratio of eigenvalues at the origin
to vary:
\begin{eqnarray*}
\dot{x} & = & -x(1+\lambda+xy) \\
\dot{y} & = & y
\end{eqnarray*}
To use \u{S}ilnikov coordinates for these systems we seek to integrate
the normal form explicitly and solve for the passage map past the
origin.  The trajectory with initial conditions $(x_0,y_0)$ is
\begin{eqnarray*}
x(t) & = & \frac{e^{-(1+\lambda)t}}{\frac{1}{x_0}+\frac{y_0}{\lambda}
(1-e^{-\lambda t})} \\
y(t) & = & y_0 e^{t}
\end{eqnarray*}
and the passage map from the line $x=1$ to the line $y=1$ has a time of
flight $t= -\ln(y)$ and an expression
\begin{equation}
x = \psi(y) = \frac{y^{(1+\lambda)}}{1+\frac{y}{\lambda}(1-y^{\lambda})}\;.
\label{pmap}
\end{equation}
When $\lambda \rightarrow 0$, this map becomes
\[ x = \psi(y) = \frac{y}{1-y \mbox{ ln}y} \;. \]
The leading order term in Equation~\ref{pmap} is $y^{(1+\lambda)}$ and
is adequate for the calculation of the unfolding of the codimension
two bifurcations of the homoclinic orbit. The global portion of the
\u{S}ilnikov approximation to the return map is an affine
transformation $y = ax+\mu$. We assume that $a \ne \pm 1$ and that
$\mu$ is the second parameter of the unfolding. Thus the approximate
return map is $\phi(y) = ay^{(1+\lambda)}+\mu$.

If the homoclinic orbit is untwisted, then $a > 0$. If the homoclinic
orbit is twisted, then $a<0$. The domain of $\phi$ is a segment with
endpoint at $y=0$. We analyze $\phi$ first in the untwisted case.
There is a periodic orbit close to the homoclinic orbit if $\phi$ has
a fixed point. The periodic orbit is degenerate if its fixed point $p$
satisfies $\phi'(p) = 1$. This happens when $\mu = p -
ap^{(1+\lambda)}$ and $a(1+\lambda)p^{\lambda} = 1$.  If we eliminate
$p$ from this pair of equations, we obtain $$\mu = \lambda
(1+\lambda)^{-(1+1/\lambda)} a^{-1/\lambda}\;.$$ As $\lambda
\rightarrow 0$, this gives $\mu \approx \lambda e^{-1}a^{-1/\lambda}$.
In order that $\mu$ be small, we need to choose the sign of $\lambda$
so that $\lambda \ln a >0$. Thus $\lambda$ is positive if $a>1$ and
$\lambda$ is negative if $0<a<1$. Observe that the curve in the
$(\lambda,\mu)$ plane along which nonhyperbolic periodic orbits occur
has a flat tangency with the curve $\mu = 0$ along which there are
homoclinic orbits. In the thin wedge between the two, there are two
periodic orbits near the location of the degenerate homoclinic orbit.
See Figure~\ref{res}.
\refstepcounter{figure} \label{res}

The analysis of the twisted case is similar, but there is indeed a new
twist. The return map has the form described above, but $a<0$. This
makes the return map $\phi$ a decreasing function of $y$. As a result,
there are two kinds of periodic orbits and two kinds of homoclinic
orbits in the unfolding: ``once rounding'' and ``twice rounding''.
The once rounding loops are twisted while the twice rounding loops are
untwisted. The once rounding periodic orbits have a negative principal
characteristic multipler while the twice rounding orbits have a
positive multiplier. The transition between the two is given by a
period doubling bifurcation. Thus there are three different types of
bifurcations that occur in the unfolding of the twisted resonant loop:
once rounding homoclinic orbits, twice rounding homoclinic orbits and
period doubling bifurcations of periodic orbits. These can be computed
in terms of the return map $\phi(y) = ay^{(1+\lambda)}+\mu$.  The once
rounding homoclinic orbits are given by $\mu = 0$. The twice rounding
homoclinic orbits are given by parameters for which $0 = \phi^2(0) =
\phi(\mu) = a \mu^{(1+\lambda)}+\mu$. Solving this equation for $\mu$
yields $\mu = |a|^{-1/\lambda}$. The period doubling bifurcations
occur at fixed points of $\phi$ for which the return map has
derivative $-1$: $a(1+\lambda)p^{\lambda} = -1$ and $\mu = p -
ap^{(1+\lambda)}$. This yields $$\mu =
(\frac{2+\lambda}{1+\lambda})(1+\lambda)^{-1/\lambda}|a|^{-1/\lambda}\;.$$
As $\lambda \rightarrow 0$, observe that the ratio of the values of
$\mu$ that produce period doubling and twice rounding homoclinic
orbits approaches $2/e$. This ratio is a ``universal'' property of
this codimension two bifurcation. As with the untwisted resonant
bifurcation, the bifurcation curves have a flat tangency.  See
Figure~\ref{restwist}.
\refstepcounter{figure} \label{restwist}

\subsection{Inclination Degeneracies}

When the principal eigenvalues of a homoclinic orbit are real, but the
homoclinic orbit is neither twisted nor untwisted, then there is an
inclination degeneracy. This situation has been studied by Deng
\cite{Deng:twisting}, and there are many subcases that lead to
different bifurcation diagrams. Here we describe one of these cases,
illustrating that horseshoes can occur in the unfolding of this
codimension two global bifurcation. The example also illustrates a
circumstance in which an approximate return map that comes from the
composition of an affine map with a local passage transformation does
not capture all of the important dynamical behavior.

Three is the lowest dimension in which one can construct a homoclinic
orbit with degenerate twist. We consider a two parameter family of
vector fields that are linear in a neighborhood of the origin, with
two real stable eigenvalues $\lambda_y < \lambda_x < 0$ and an
unstable real eigenvalue $\lambda_z > 0$. We assume that there is a
homoclinic orbit that approaches the origin along the $x$ axis in
forward time and along the $z$ axis in backward time. We shall assume
further that the return map for the orbit is degenerate in the
twisting of the normal bundle around the homoclinic orbit. This means
that a vector, pointing in the direction of the principal stable
eigenvector as the homoclinic orbit leaves the origin, returns in the
direction of the strong stable manifold (rather than the generic
behavior of returning in the direction of the unstable manifold).

Let us view the situation in terms of a \u{S}ilnikov approximation.
The passage map near the origin from the cross section $\Sigma$
defined by $x=1$ to the cross section $z=1$ will be given by
\[
   \left( \begin{array}{c} u\\v \end{array} \right) =
   \psi \left( \begin{array}{c} y \\ z \end{array} \right) =
   \left( \begin{array}{c} z^{\alpha} \\ y z^{\beta} \end{array} \right)
\]
where
\[
  \alpha = -\lambda_x/\lambda_z,\; \beta = -\lambda_y/\lambda_z,\;
  0<\alpha <\beta \; .
\]
Here and below, if $z<0$, we denote $ z^{\alpha} = -(-z)^{\alpha}$.
The global return from $z=1$ to $x=1$ will be approximated by
\[
  \left( \begin{array}{c} y\\z \end{array} \right) =
  \gamma \left( \begin{array}{c} u\\v \end{array} \right) =
  \left( \begin{array}{c} c \\ \lambda \end{array} \right) +
  \left( \begin{array}{c} au+bv \\ -\mu u + u^2 + v \end{array} \right) \;\; .
\]
Then the return map $\phi$ is approximated by
\[
  \phi \left( \begin{array}{c} y\\z \end{array} \right) =
  \left( \begin{array}{c} c + a z^{\alpha} + b y z^{\beta} \\
  \lambda - \mu z^{\alpha} + z^{2\alpha} + y z^{\beta}
  \end{array} \right) \;\; .
\]
Here the parameter $\lambda$ determines whether a homoclinic
connection occurs, while the sign of $\mu$ determines whether the
connection is twisted or untwisted. Note that the approximation has
included a quadratic term in $\gamma$. Without this term, the plane
spanned by the principal eigenvectors would return entirely within the
stable manifold, a highly degenerate situation.

There are now several cases determined by the relative magnitude of
the exponents $\alpha$ and $\beta$. The case we shall examine is the
one in which we assume that $2\alpha < \beta$ and $\alpha < 1/2$.  The
first of these assumptions implies that $|y z^{\beta}| =
o(z^{2\alpha})$ and the second leads to complex dynamical behavior
involving the stretching in the $z$ direction. We shall not try to
give a complete analysis of even the specific family of maps defined
by $\phi$, but we shall show that there are horseshoes that appear in
some regions of the parameter space for this family.

The assumption $2\alpha < \beta$ leads to an approximation of $\phi$
by a one-dimensional mapping. The nature of the approximation will be
examined below. Consider the one-dimensional mapping $\hat{\phi}(z) =
\lambda - \mu z^{\alpha} + z^{2\alpha}$ obtained by ignoring the $y$
coordinate in the definition of $\phi$.  This is a unimodal map, and
we look for ranges of parameter values for which this mapping will
have a hyperbolic invariant set. If $\lambda = 3\mu^2/16$ and $\mu
>0$, $\hat{\phi}(z) = (z^{\alpha} - \mu/4)(z^{\alpha}- 3\mu/4)$.  Let
\[
	A=0,\; B=(\mu/4)^{1/\alpha},\; C=(3\mu/4)^{1/\alpha},\;
D=\mu^{1/\alpha}\; ,
\]
so $A<B<C<D$.  Then $\hat{\phi}(A) = \hat{\phi}(D) = 3\mu^2/16 \gg D$
since $\alpha<1/2$, and $\hat{\phi}(B) = \hat{\phi}(C) = A$.  Now
$|\hat{\phi}'(B)| = |\hat{\phi}'(C)| = 2 \alpha (-\mu/4)^{2-1/\alpha}
> 1$ for $-\mu$ small.  Since $\hat{\phi}([A,B]) \supset [A,D],\;
\hat{\phi}([C,D])
\supset[A,D]$, and $\hat{\phi}>1$ on $[A,B]$ and $[C,D]$, we conclude that
the map $\hat{\phi}$ has a hyperbolic invariant set in the interval
$[A,D) = [0,\mu^{1/\alpha})$.

In the cross section $\Sigma$, consider the rectangle $R_\mu =
[c-2\mu^{1/\alpha},c+2\mu^{1/\alpha}] \times
[-2\mu^{1/\alpha},2\mu^{1/\alpha}]$. If we set $\lambda = 3\mu^2/16$
and let $\mu \rightarrow 0$ from below, then the return maps of the
rectangles $R_\mu$ can be rescaled so that they tend to a map of rank
1.  The rescaled maps have a vertical coordinate that behaves as a
rescaled version of $\hat{\phi}$. This situation is analogous to the
behavior of the H\'{e}non map \cite{Henon} in the limit as the map
tends to a map of rank one. The hyperbolic invariant set will persist
as a horseshoe as we leave the singular limit $\mu \rightarrow 0$.
The analysis of Benedicks and Carleson \cite{BenedicksCarleson},
extended by Mora and Viana \cite{MV} indicates further that as
$\lambda$ varies, we will expect to encounter chaotic attractors for
the return map $\phi$.

\subsection{Multiple Connections}

Consider a vector field in a generic two-parameter family that has an
equilibrium point with a single unstable eigenvalue, real non-resonant
principal eigenvalues and a pair of generic homoclinic orbits formed
from the two unstable separatrices of the saddle point.  We would like
to analyze the unfoldings of such systems.  In the planar case, this
determines the geometry of the unfolding of the family and is
described below. On the other hand, there are several choices that
occur if the vector field has dimension at least three. In particular,
if the two homoclinic orbits approach the equilibrium from the same
direction along the principal stable eigenvector, then there are cases
that lead to either geometric Lorenz attractors \cite{GW} or to
complex periodic orbits formed by the {\em gluing bifurcations}
described by Gambaudo, Glendinning and Tresser \cite{GGT}. We recall
bits of their analysis below.

Let $X$ be a planar vector field with a non-resonant hyperbolic
equilibrium point at the origin and a pair of orbits homoclinic to the
origin. Assume that $X$ is contained in a generic two parameter family
of vector fields.  There are two topologically distinct configurations
depending upon whether one homoclinic orbit is contained in the other,
but there is little difference in the analysis of the two cases.  Pick
a pair of cross sections $\Sigma_1, \Sigma_2$ to the two stable
separatrices and consider the return map $\phi : \Sigma_1 \cup
\Sigma_2 \rightarrow \Sigma_1 \cup \Sigma_2$.  This return map will
have discontinuities at the intersections of $\Sigma_1$ and $\Sigma_2$
with the stable manifold of $0$. Thus we can think of $\phi$ as
constructed from four maps $\phi_{ij} : \Sigma_i \rightarrow
\Sigma_j$.  These maps fit together so that the ranges of $\phi_{1j}$
and $\phi_{2j}$ are contiguous and do not overlap. At the common
endpoint of the images, there is a singularity of the form
$x^{\alpha}$ with $\alpha \ne 1$. By reversing time if necessary, we
may assume that the magnitude of the stable eigenvalue of $0$ is
larger than the unstable eigenvalue, implying $\alpha > 1$ and that
the derivatives of the $\phi_{ij}$ are smaller than $1$ near the
stable manifold of $0$.  Furthermore, we may choose orientations of
$\Sigma_1$ and $\Sigma_2$ so that the maps $\phi_{ij}$ are increasing.
Then each $\phi_{ij}$ or iterates of the $\phi_{ij}$ in a connected
domain can have at most one fixed point, and that point will be
stable.  Determining which types of periodic orbits bifurcate from the
double loop becomes a matter of determining which configurations of
fixed points for iterates of the maps $\phi_{ij}$ can occur.  The
cross sections and return maps are pictured in Figure \ref{gluing}.
\refstepcounter{figure} \label{gluing}

Denote once again $x^{\alpha} = -(-x)^{\alpha}$ for $x<0$ and choose
coordinates on cross sections that are centered on the stable
manifolds.  The \u{S}ilnikov approximation for the return maps have
the form $\phi_{ij}(x) = a_jx^{\alpha} + b_j$, but one must remember
that there are two components to the domain of the return map and
$\phi_{12}$ and $\phi_{21}$ map one domain to the other. Here $b_1$
and $b_2$ can be regarded as the unfolding parameters of the system.

It is important here that the maps $\phi_{1j}$ and $\phi_{2j}$ have
contiguous images. There are three kinds of periodic orbits for the
return maps that correspond to periodic orbits of the flow: fixed
points of $\phi_{ii}$ and fixed points of a composition $\phi_{12}
\circ \phi_{21}$ or $\phi_{21} \circ \phi_{12}$.  The fixed points of
$\phi_{ii}$ correspond to periodic orbits that lie close to one of the
homoclinic orbits, while the fixed points of $\phi_{12} \circ
\phi_{21}$ and $\phi_{21} \circ \phi_{12}$ correspond to periodic
orbits that lie close to both of the homoclinic orbits from the
system. All of the periodic orbits in a neighborhood of the double
homoclinic orbit have a stability that is determined by whether the
Jacobian of the non-resonant saddle at the origin has a positive trace
(unstable) or negative trace (stable).  The boundaries between the
parameter regions with different types of periodic orbits will be
given by parameter values at which homoclinic orbits occur. There are
four types of homoclinic orbits. In terms of the return map these
correspond to fixed points at $0$ of $\phi_{11}$, $\phi_{22}$,
$\phi_{12} \circ \phi_{21}$ and $\phi_{21} \circ \phi_{12}$.  The
parameter curves yielding the first two types of homoclinic orbits are
simply $b_1 = 0$ and $b_2 = 0$. The parameter values yielding the more
complicated homoclinic orbits are given by $a_2 b_1^{\alpha} + b_2 =
0$ and $a_1 b_2^{\alpha} + b_1 = 0$. Since the return maps are
orientation preserving, $a>0$, and taking the domain and ranges into
account, these curves of bifurcating homoclinic orbits occur in the
quadrants of the $(b_1,b_2)$ plane in which $b_1$ and $b_2$ have
opposite signs. This completes the description of the unfolding of
this codimension two bifurcation.

In higher dimensions, double homoclinic orbits can be more complicated
than the ones that occur for planar vector fields. The additional
complication is due to the fact that the return maps along two
separatrices of a one-dimensional unstable manifold need not match in
the same fashion that they do for a planar vector field. If the
unstable eigenvalue is smaller in magnitude than all of the stable
eigenvalues, then the periodic orbits that bifurcate from the
homoclinic cycles are all stable, and there are only a finite number
of them for any given parameter value. Indeed there are at most two.
The situation that we describe was analyzed by Gambaudo, Glendinning
and Tresser \cite{GGT}, who called this the {\em gluing bifurcation}.
Their analysis is based upon the study of return maps for the double
cycle. Let us describe a bit more.

Let $X$ be a vector field on $R^n$ with an equilibrium point at the
origin having a single unstable eigenvalue $\lambda_u$ and a stable
eigenspace of dimension $n-1$ whose spectrum lies to the left of the
line $Re(\lambda) < -\lambda_u$. Assume further that both unstable
separatrices of the origin are homoclinic, and that $X$ is embedded in
a generic two parameter family. Form a cross section $\Sigma$ to the
stable manifold $W^s(0)$ near the origin and let the components of
$\Sigma - W^s(0)$ be $\Sigma_1$ and $\Sigma_2$. The return maps
$\phi_1$ and $\phi_2$ of $\Sigma_1$ and $\Sigma_2$ will be continuous,
contracting and have images that are punctured neighborhoods of the
(first) intersection of each unstable separatrix with $\Sigma$. Thus
we can abstract the situation that we encounter to the study of the
``quasicontractions'' $\phi_1: \Sigma_1 \rightarrow \Sigma $ and
$\phi_2: \Sigma_2 \rightarrow \Sigma$. If the images of these maps
intersect $W^s(0)$, then there will be more components formed from
subsequent compositions of $\phi_1$ and $\phi_2$. Using ``symbolic
dynamics'', one can analyze the periodic orbits that can form from the
iterated compositions of a pair of maps. The following one-dimensional
model is adequate to represent the general situation. Consider a map
$f:[-1,1] \rightarrow [-1,1]$ that is discontinuous at the origin and
contracting. The ``kneading theory'' or symbolic dynamics of $f$
characterize its behavior. This gives a combinatorial procedure for
determining the dynamics of $f$ from the trajectories of the points
$-1,0^-,0^+,1$. For maps of the form of $f$, there are zero, one or
two periodic orbits. The patterns of signs of $f^i(x)$ along periodic
orbits are greatly restricted and can be assigned a {\em signature}
that is a rational number. When there are two periodic orbits, their
signatures $p/q, p'/q'$ are {\em Farey neighbors}; i.e.,
$|pq'-qp'|=1$. The different families of periodic orbits appear and
disappear in complex homoclinic orbits.  Thus, even without the
occurrence of horseshoes, the gluing bifurcation has an unfolding with
an infinite number of curves in its bifurcation set.

Other connecting orbit bifurcations have been studied by
Chow, Deng and Fiedler \cite{ChowDeng:bifurcation},
Deng \cite{Deng:countable}, Glendinning and Sparrow
\cite{GlendinningSparrow:tpoints}, Glendinning and Tresser
\cite{GlendinningTresser:hyperchaos}, and
Schecter \cite{Schecter:simultaneous,Schecter:singularity}

\subsection{Computations of Connecting Orbits}

The numerical computation of connecting orbits has only recently been
investigated, and the construction of robust algorithms that
effectively compute connecting orbits in large classes of vector
fields remains a challenge. We make a few comments concerning the
numerical difficulties and fewer suggestions for how these
difficulties might be confronted. We seek to solve the following
problems. In generic one parameter families of vector fields, there
are isolated points with connecting orbits between equilibrium points.
We want accurate calculations of these parameter values and of the
resulting orbits.  (There may also be accumulation points of
connecting orbits of increasing length and complicated topological
structure in generic one parameter families
\cite{BaesensGuckenheimer:three}.) In generic two parameter families
of vector fields, there are curves of parameter values at which
connecting orbits occur. We seek to use continuation methods to
compute these curves. The endpoints of these curves of homoclinic and
heteroclinic bifurcations are frequently global codimension two
bifurcations. We seek to classify these codimension two bifurcations
and construct algorithms for their computation. The unfoldings of some
of these codimension two bifurcations have been described above, but
there is a long list of cases that have yet to be analyzed fully --
even at the level of \u{S}ilnikov approximations.  Thus the goal of
implementing the computation of unfoldings of global codimension two
bifurcations involves mathematical as well as computational questions.

Generic connecting orbits are intersections of stable and unstable
manifolds of equilibria. Therefore, algorithms for the reliable
calculation of these invariant manifolds might seem to form the basis
for computation of the connecting orbits. This strategy is clearly an
expensive one from a computational point of view, so we seek more
direct methods for computing connecting orbits. In practice, most
examples in higher dimensional vector fields have been computed by
tracking periodic orbits that bifurcate from homoclinic orbits, with
high period periodic orbits used as approximations to the homoclinic
orbits. When the periodic orbits involved are not stable, then
tracking the periodic orbits requires the use of algorithms for
solving boundary value problems or algorithms that find fixed points
of a cross section. Since the periodic orbits usually disappear at the
homoclinic orbit, these computations are hard to implement in an
automatic fashion. We seek methods that are more direct.

Mathematically, a connecting orbit is the solution of a boundary value
problem on an infinite interval of time. To construct algorithms based
upon boundary value solvers for finding connecting orbits, one wants
to convert the problem into one involving a finite time interval. This
can be done approximately by introducing linear (or polynomial)
approximations for the local stable and unstable manifolds containing
the ends of the connecting orbits and seeking trajectories that begin
on the local unstable manifold and end on the local stable manifold.
Beyn \cite{Beyn}, Chow and Lin \cite{CL}, Doedel and Friedman
\cite{Friedman,DF} and Schecter \cite{Schecter2} have all considered
algorithmic aspects of the computation of homoclinic orbits, but the
only computations that have been examined carefully involve planar
vector fields.

One of the pragmatic questions concerning the computation of
homoclinic bifurcations is whether the goal is to obtain an accurate
approximation of the parameter values at which the bifurcation occurs
or whether one is primarily interested in the computation of an
accurate approximation to the homoclinic orbit. These are
substantially different questions for reasons that we now describe.
Suppose a small error has been made in the determination of the
parameter value $\mu$ for which there is a homoclinic orbit in a
system. We ask whether there is a trajectory for parameter value $\mu
'$ that closely approximates the homoclinic orbit that exists for
parameter value $\mu$.  The distance of closest approach between the
stable and unstable manifolds will be of order $\mu - \mu '$ since
compact portions of the manifolds vary smoothly with the parameters.
Thus we want to know how close the trajectory through a point close to
the stable manifold comes to the equilibrium. Estimates can be
obtained from linear vector fields.

Consider a linear vector field $X$ with a hyperbolic equilibrium point
at the origin and assume that the stable and unstable manifolds are
coordinate subspaces. The vector field ``separates'' into a stable
system and an unstable system. Along trajectories, the unstable
coordinates grow at an exponential rate. Now look at a generic point
$x$ on the stable manifold and estimate the distance of closest
approach to the equilibrium point of a trajectory passing through a
point $x'$ close to $x$. If the difference $x'-x$ has a non-zero
component in the direction of the strongest unstable eigenvalue of
$X$, then this component will grow exponentially at a rate which is
the eigenvalue of the strongest unstable eigenvalue. As we saw above,
a generic point on the stable manifold approaches the equilibrium in
the direction of the principal (weakest) stable eigenvalue. Putting
these observations together, we estimate the point of closest approach
to the origin of the trajectory through $x'$ as having order
$c^{\beta}$ where $c$ is the component of $x'$ in the strongest
unstable direction and $\beta$ is the ratio of the magnitude of the
principal stable eigenvalue to the strongest unstable eigenvalue. When
$1/\beta$ is large, then small errors in locating a point on the
unstable eigenvalue lead to much larger distances between the closest
trajectory and the homoclinic orbit that is sought. If we formulate a
boundary value problem to find the homoclinic orbit, then we will need
to start with a very good approximation to the homoclinic orbit to
have hope of being able to use the boundary value solver.
Trajectories depend very sensitively upon the parameters in a
neighborhood of the equilibrium. These difficulties become still more
extreme if one seeks the computation of a unifocal homoclinic orbit.

We have attempted to point out some of the difficulties associated
with the computation of connecting orbits.  The computation of these
dynamical structures is important due to their role in organizing the
dynamics of a system.  Consequently, this is an interesting problem.

\section{An Example: The Hodgkin and Huxley Equations}

This section describes the analysis of a moderately complicated
dynamical system and is work done in collaboration with Isabel
Labouriau.  The Hodgkin and Huxley (HH) equations are a ``simple''
neuron model developed from experiments performed with a squid
\cite{HH52}.  These equations relate the difference of electric
potential across the cell membrane ($V$) and gating variables ($m$,
$n$, and $h$) for ion channels, to the stimulus intensity ($I$), and
temperature ($T$), as follows:
\[
\left\{ \begin{array}{lcl}
\dot V &=&-G(V,m,n,h) + I\\
\dot m &=&\Phi (T) \left[(1-m)  \alpha _m(V) - m \beta _m(V)\right]\\
\dot n &=&\Phi (T) \left[(1-n)  \alpha _n(V) - n \beta _n(V)\right]\\
\dot h &=&\Phi (T) \left[(1-h)  \alpha _h(V) - h \beta _h(V)\right]
\end{array}\right. \;\;\;\;\; \mbox{(HH)}
\]
where $\dot x$ stands for $dx/dt$ and $\Phi$ is given by
$\Phi (T)=3^{(T-6.3) / 10}$. The other functions involved are:
\[
G(V,m,n,h)=\bar{g}_{\rm Na} m^3h(V-\bar{V}_{\rm Na}) + \bar{g}_{\rm K}
n^4(V-\bar{V}_{\rm K}) + \bar{g}_{\rm L}(V-\bar{V}_{\rm L})
\]
and the equations modeling the variation of membrane permeability:
\[
\begin{array}{rclrcl}
\alpha _m (V)&=&\Psi ({{V+25} \over 10})
&\beta _m (V)&=&4 e^{V /18}
\\
\alpha _n (V)&=&0.1\Psi ({{V +10}\over 10})
&\beta _n (V)&=&0.125 e^{V/{80}}
\\
\alpha _h (V)&=&0.07e^{V / 20}
&\beta _h (V)&=&\left( 1+e^{(V+30)/ 10}\right)^{-1}
\end{array}
\]
\[
{\rm with}\qquad \Psi (x)=\left\{ \begin{array}{ll}
x /(e^x-1) & {\rm if } \  x \ne 0\\
1& {\rm if} \  x= 0
\end{array} \right. \;\; .
\]
Notice that $\alpha _y(V) + \beta _y(V) \ne 0 $ for all $V$ and for
$y=m$, $n$ or $h$. The parameters $\bar g_{\rm ion}$, $\bar{V}_{\rm
ion}$ representing maximum conductance and equilibrium potential for
the ion were obtained from experimental data by Hodgkin and Huxley,
with the values given below:
\[
\begin{array}{lll}
\bar{g}_{\rm Na}= 120\; {\rm mS}/{\rm cm}^2
&\bar{g}_{\rm K}= 36\;{\rm mS}/ {\rm cm}^2
&\bar{g}_{\rm L}= 0.3\; {\rm mS}/{\rm cm}^2\\
\bar{V}_{\rm Na}= -115\; {\rm mV}
&\bar{V}_{\rm K}= 12\; {\rm mV}
&\bar{V}_{\rm L}=10.599\; {\rm  mV}
\end{array}
\]
The values $\bar{V}_{\rm Na}$ and ${\bar{V}_{\rm K}}$ can be controlled
experimentally
\cite{HH52b}. For the results in this paper, we use the temperature
$T=6.3^\circ$C and, except where stated explicitly, all the other
parameters involved in the HH equations have the values quoted above
that we call the HH values.

We describe some of the bifurcations of the HH equations as an
illustration of the theory and algorithms described above.  The local
bifurcations of equilibria were calculated in terms of the the
derivatives of the HH equations at the equilibrium points.  Global
bifurcations could only be studied by numerically integrating the HH
equations.  Our analysis of local bifurcations used the symbolic
computer program Maple to implement the calculation of saddle-node and
Hopf bifurcation curves.  For $y=m,n,$ or $h$ the equation for $\dot
y$ in (HH) is linear in $y$, so the last three components of an
equilibrium solution $(V_\ast,M_\ast,N_\ast,H_\ast)$ of (HH) can be
written as functions of $V_\ast$: $$y_\ast = y_\infty(V_\ast) =
{{\alpha_y(V_\ast)} \over {\alpha_y(V_\ast)+\beta _y(V_\ast)}} \qquad
{\rm for\ } y=m, n, h \; .$$ Substituting $y_\ast$ for $y=m, n, h$ in
the first equation, we get:
\[
G(V_\ast,m_\infty(V_\ast),n_\infty(V_\ast),h_\infty(V_\ast) )
= f(V_\ast) = I \; .
\]
Thus, for fixed ${\bar{V}_{\rm K}}$ there is exactly one value of $I$
for which $ (V_\ast,m_\ast,n_\ast,h_\ast )$ is at equilibrium. Note
that derivatives of (HH) are independent of $I$.

When ${\bar{V}_{\rm K}}$ has the HH value of $12$ mV, $f$ is monotonic
and (HH) has a unique equilibrium for each value of $I$. For fixed
lower values of ${\bar{V}_{\rm K}}$, there are two saddle-node
bifurcations as $I$ is varied, creating a region with three
equilibria. The two curves of saddle-nodes terminate at a cusp point.
See also \cite{holden}. The saddle-node curves in the $I \times
{\bar{V}_{\rm K}}$ plane were computed parametrically with $V_\ast$ as
the independent parameter.  The equations describing the saddle-node
curves involve the determinant of the matrix of first derivatives of
(HH) at an equilibrium point.  We calculated an explicit expression
for this determinant symbolically with Maple.  By solving the equation
that the determinant vanishes for ${\bar{V}_{\rm K}}$ at equilibrium
values of $(V_\ast,m_\ast,n_\ast,h_\ast )$, we obtained the curve
${\bar{V}_{\rm K}} (V_\ast)$ of parameter values corresponding to zero
eigenvalues.

To determine the parameter values at which Hopf bifurcation occurs, it
is necessary to compute eigenvalues of the matrix of first derivatives
of (HH) at an equilibrium point.  There is a pair of purely imaginary
eigenvalue when the characteristic polynomial $x^4+ c_3 x^3 + c_2 x^2
+ c_1 x + c_0$ of this matrix satisfies simultaneously the third
degree equation $c_1^2 - c_1 c_2 c_3 + c_0 c_3^2 = 0$ and the
inequality $c_1 c_3 > 0$. These are the expressions that result from
the Sylvester resultant calculation described earlier.  Again, we
computed this equation symbolically, assuming a given value of
$V_\ast$, and solved for ${\bar{V}_{\rm K}}$.  The graph we obtained
for the solution of this equation and inequality disagrees slightly
with the findings of Holden, {\em et al.}, \cite{holden} for the HH
value ${\bar{g}_{\rm K}} = 36$.  Takens-Bogdanov bifurcations occur
when the equations defining Hopf bifurcations and saddle-node
bifurcations are satisfied simultaneously.

The saddle-node and Hopf bifurcations are the only codimension one
bifurcations that can be computed explicitly from (HH) without
numerical integration.  The presence of double cycles where the two
periodic orbits created at Hopf bifurcation points coalesce and
disappear has been established previously \cite{Hass2,Lab,R&M} and the
existence of saddle loops emanating from the Takens-Bogdanov points is
predicted by bifurcation theory \cite{GH}. To determine further
information about global bifurcations, we rely upon numerical
integrations that were performed with the computer program {\em
DsTool} \cite{dstool:AMS}. This program establishes a graphical
interface and display for investigating bifurcations of dynamical
systems. It allows one to mark points in a two-dimensional parameter
space with identifying symbols and to display phase portraits that
correspond to these points. The computed data for local bifurcations
was displayed in the parameter space window.  By searching for the
boundaries of parameter regions yielding structurally stable dynamics
and using our knowledge of the unfoldings of codimension two
bifurcations, we deduced the location of curves along which global
bifurcations take place. We obtain a consistent picture of the
bifurcation diagrams for (HH) in the two-dimensional
$I\times{\bar{V}_{\rm K}}$ parameter plane. These diagrams have not
been proved to be correct, but they are based upon strong numerical
evidence.

The bifurcation diagram resulting from our numerical investigations is
shown in Figure~\ref{hh1}.
\refstepcounter{figure} \label{hh1}
Its main features include a curve of
double cycles (dc) which enters the cusp region with three equilibrium
points and terminates at a degenerate Hopf bifurcation (dh) close to
the Takens-Bogdanov point (tb). These double cycles are the ones
described in \cite{Lab}.  The curve of saddle loops (sl) emanating
from the Takens-Bogdanov point crosses the Hopf curve beyond the
degenerate Hopf point, and then turns sharply. From this sharp bend,
it proceeds almost parallel to the saddle-node curve (sn). The saddle
loops appear to undergo a reversal of orientation along this portion
of the curve (sl). After the orientation has reversed, one encounters
a set of parameters at which the unstable eigenvalue and the weakest
stable eigenvalue have equal magnitudes. This point (tsl) in parameter
space is a twisted neutral saddle loop, and there are additional
curves of untwisted twice rounding and period doubling bifurcations
that emerge from (tsl).

When ${\bar{g}_{\rm K}}$ is decreased from the HH value of $36\; {\rm
mS}/{\rm cm}^2$ the Takens-Bogdanov point in the $I\times{\bar{V}_{\rm
K}}$ plane moves towards the cusp point and past it. This agrees
qualitatively with the findings of \cite{holden}, but their results
differ from ours in the value of ${\bar{g}_{\rm K}}$ for which the
Takens-Bogdanov point moves past the cusp. The unfolding for the
codimension three bifurcation in which cusp and Takens-Bogdanov
bifurcations coincide has been analyzed by Guckenheimer \cite{GPD} and
Dumortier and Roussarie \cite{Dumortier}. The geometry of the
unfolding of this codimension three bifurcation can be visualized by
drawing a two dimensional sphere that encloses the codimension three
point in the three dimensional parameter space of the unfolding
\cite{GPD}.

To further explore the effect of this codimension three bifurcation on
the bifurcation diagrams of (HH), we decreased ${\bar{g}_{\rm K}}$
from the HH value of $36\; {\rm mS}/{\rm cm}^2$ to $12\; {\rm mS}/{\rm
cm}^2$ and computed another bifurcation diagram in the
$I\times{\bar{V}_{\rm K}}$ plane.  The new bifurcation diagram is
shown in Figure \ref{hh2}.
\refstepcounter{figure} \label{hh2}
Among its features are a curve of double
cycles (dc) that terminates at a neutral saddle loop point (nsl)
instead of a double Hopf bifurcation as in the unfolding of the
codimension three bifurcation.  The point (nsl) does not lie on the
saddle loop branch emanating from the Takens-Bogdanov point (tb),
however. Instead it ends on a saddle loop that encloses both
equilibrium points.  This branch of saddle loops ends on both branches
of saddle-nodes at saddle node loops (snl). The branch of twisted
saddle loops (tsl) that was present at the higher value of
${\bar{g}_{\rm K}}$ remains. It starts on the saddle-node curve at
another saddle node loop. The twisted saddle loops still passes
through a neutral point (tnsl) at which bifurcation curves of period
doublings (pd) and doubled saddle loops (dsl) originate.  Our proposed
bifurcation diagrams for (HH) near the cusp points appear to be
compatible with the unfolding of the Takens-Bogdanov cusp codimension
three bifurcation, though the diagrams drawn here are sufficiently far
from the codimension three bifurcation that significant differences
with its unfolding exist.

The intent of this rapid tour of the complicated dynamics of a
realistic biological model was to draw attention to the interplay
between theory and numerical exploration.  Since many regions in the
parameter space are very small, the theory helps to guide the
exploration to discover the correct bifurcation diagram.

\section{The Effects of Symmetry}

In the previous sections, non-generic bifurcation has arisen in
multiparameter families of vector fields and is a codimension two, or
higher, phenomenon.  Non-generic bifurcation is often encountered in
the real world as a result of symmetry and in this section we discuss
the bifurcation of symmetric vector fields.

Within a generic one parameter family of vector fields,
$\dot{x}=f_\lambda(x)$, bifurcations of equilibria are either
saddle-node bifurcations, when a single eigenvalue passes through 0,
or Hopf bifurcations, when a pair of complex conjugate eigenvalues
cross the imaginary axis.  These are both well understood \cite{GH},
but with symmetry, generic bifurcation can be much more complicated.
One of our goals has been to explore how complex local bifurcation
with symmetry can be.

We first make a few definitions.  Let $G$ be a finite subgroup of
$O(n)$ acting on $R^n$.  An equation $\dot{x}=f(x)$ is {\em
$G$-equivariant} or {\em symmetric} if $$f(gx)=g \circ f(x) \mbox{ for
all } g \in G \; .$$ For a fixed group $G$, we wish to study generic
bifurcation within the class of $G$-equivariant vector fields.

We now make two assumptions on the group action. The first is that the
action of $G$ is absolutely irreducible.  This means that the only
matrices which commute with $G$ are scalar multiples of the identity.
The second is that the action of $G$ is fixed point free: the only
point fixed by $G$ is the origin.  These assumptions have two
consequences for $G$-equivariant vector fields.  The first is that the
origin is always an equilibrium point of a $G$-equivariant vector
field, since $$ f(0)=f(g0)=gf(0) \Rightarrow f(0)=0\; . $$ The second
consequence is that the linearization at the origin is a multiple of
the identity.  This follows from the calculations:
\[ \begin{array}{l} D(f \circ g)|_0=Df|_{g(0)} \; g=Df|_0 \; g \\
   \mbox{and } D(f \circ g)|_0=D(g \; f)|_0=g \; Df|_0 \end{array} \]
Because $Df|_0$ commutes with all $g\in G$ and $G$ is absolutely
irreducible, $Df|_0$ must be a scalar multiple of the identity.  We
may now state the following:

\begin{theorem}
Let $G$ be fixed point free and absolutely irreducible and let
$\dot{x}=f_\lambda(x)$ be a one-parameter family of $G$-equivariant
vector fields.  If the origin undergoes a bifurcation with one zero
eigenvalue for $\lambda=\lambda_0$, then the linearization at the
origin is given by $Df_\lambda |_{(0,\lambda_0)}=0$.  In other words,
all the eigenvalues must pass through zero simultaneously.
\end{theorem}
This degenerate linearization occurs as a codimension one phenomenon
due to the symmetry of the problem.  We wish to analyze the
bifurcations at the origin, but the possibilities are not clear and
there are still many restrictions due to the equivariance.

To study the bifurcation in an analogous fashion as for the
non-symmetric case, we will compute a normal form by computing a
Taylor series at the origin: $$
f_{\lambda}(x)=\sum_{i=1}^{\infty}P_i(x) \; .$$ Each $P_i$ is a
homogeneous, degree $i$, $G$-equivariant.  The $P_i$ are not always
easy to calculate and we are led to the study of polynomial invariants
and equivariants which combines elements of combinatorial theory and
modern commutative algebra~\cite{GSS,Stanley:invariants}, but we shall
not delve into this topic in this paper.

We shall now study a simple example: $D_4$ acting on $R^2$, which may
be thought of as the symmetry of the square in the plane.  Generators
for this action are given by
\[
\left( \begin{array}{cc} 1 & 0 \\ 0 & -1 \end{array} \right) \mbox{ and }
\left( \begin{array}{cc} 0 & 1 \\ 1 & 0 \end{array} \right)\;,
\]
and it is easy to show that this action is fixed point free and
absolutely irreducible.  We look at the degree three power series
expansion at the origin:
\begin{eqnarray*}
	\dot{x} & = & \lambda x + a x^3 +b x y^2 \\
	\dot{y} & = & \lambda y + a y^3+b x^2 y
\end{eqnarray*}
As $\lambda$ passes through 0, the origin changes stability, but what
else happens in the process?  We can get some additional information
from looking at subgroups of $G$ and fixed point subspaces.

If $H\subset G$, then we define the fixed point subspace of the
subgroup $H$ by $$\mbox{Fix}(H)=\{x\in R^n\mbox{ such that }hx=x\mbox{
for all }h\in H\}.$$ Fixed point subspace are important because they
are invariant under the flow.  This is easy to see because if $x\in
\mbox{Fix}(H)$, then $f(x)=f(hx)=hf(x)$ so $f(x)\in\mbox{Fix}(H)$,
{\em i.e.}, the vector field on $\mbox{Fix}(H)$ is tangent to
$\mbox{Fix}(H)$.  The fixed point subspaces for the example are the
axes of symmetry plus the trivial ones: the origin and all of $R^2$.

To explain bifurcation in some of the fixed point subspaces, Cicogna
and Vanderbauwhede formulated the following lemma:
\begin{lemma} (The Fixed Point Branching Lemma)
\cite{GSS} If $V$ is a one-dimensional fixed point subspace,
then there is a branch of solutions in $V$ in $R^n \times R$ that pass
through $(0,0)$ and is part of a generic bifurcation.
\end{lemma}
For the $D_4$ example, this means that on the axes of symmetry there
are pitchfork bifurcations.  For this simple problem, we may verify
this result by explicitly computing the equilibrium points which are
given by:
\[(0,0)\]
\[(\pm\sqrt{\frac{-\lambda}{a}},0),(0,\pm\sqrt{\frac{-\lambda}{a}})\]
\[(\pm\sqrt{\frac{-\lambda}{a+b}},\pm\sqrt{\frac{-\lambda}{a+b}})\]
So if $a<0$ and $a+b<0$, then as $\lambda$ passes through zero, eight
new equilibrium points are created.  It is typical behavior to get
multiple equilibrium points in equivariant bifurcations and a
substantial theory has been developed by M. Field
\cite{Field:equivariant2}, M. Golubitsky \cite{GSS} and others.  This
work, based on group theory and singularity theory, can give
information about bifurcation to equilibrium points, period orbits and
quasiperiodic orbits.  However, this is a far from complete
categorization of the behavior which may be obtained from more
complicated group actions.

Now consider the action of a group $G$ on $R^3$ generated by
$$
\left( \begin{array}{ccc} -1 & 0 & 0 \\ 0 & 1 & 0 \\ 0 & 0 & 1 \end{array}
\right)
\mbox{ and }
\left( \begin{array}{ccc} 0 & 1 & 0 \\ 0 & 0 & 1 \\ 1 & 0 & 0 \end{array}
\right) \; .
$$ The action is fixed point free, absolutely irreducible and, a few
short calculations show, the degree three equivariant vector field has
the form:
\begin{eqnarray}
  \dot{x}_1 & = & (\lambda+a_1x_1^2+a_2x_2^2+a_3x_3^2)  x_1 \nonumber \\
  \dot{x}_2 & = & (\lambda+a_1x_2^2+a_2x_3^2+a_3x_1^2)  x_2 \label{GH} \\
  \dot{x}_3 & = & (\lambda+a_1x_3^2+a_2x_1^2+a_3x_2^2)  x_3 \nonumber
\end{eqnarray}
One of the consequences of the symmetry is the invariance of both the
coordinate axes and the coordinate planes.  Guckenheimer and Holmes
\cite{GuckenheimerHolmes:heteroclinic} observed that this system of
equations has structurally stable heteroclinic cycles for open regions
of the parameter space.

\begin{theorem}
If either $a_3<a_1<a_2<0$ or $a_2<a_1<a_3<0$ and $2a_1>a_2+a_3$ then
Equations~\ref{GH} have attracting heteroclinic cycles for $\lambda>0$
and a globally attracting equilibrium at the origin for $\lambda \leq
0$.
\end{theorem}
This theorem essentially follows from examining the flow in the
invariant coordinate planes.  The consequence is that within families
of $G$-equivariant vector fields there are generic one parameter
bifurcations from an attracting equilibrium point to structurally
stable attracting limit cycles.  We continue to study more complicated
symmetry groups in an attempt to discover how complicated post
bifurcation behavior can be.  The answer is that a stable equilibrium
point can bifurcate directly to a chaotic attractor of small amplitude
\cite{GuckenheimerWorfolk:instant}.

Consider the action of $\bar{G}$ on $R^4$ generated by
\[
\left( \begin{array}{cccc} -1 & 0 & 0 & 0 \\ 0 & 1 & 0 & 0 \\
			0 & 0 & 1 & 0 \\ 0 & 0 & 0 & 1 \end{array} \right)
\mbox{ and }
\left( \begin{array}{cccc} 0 & 1 & 0 & 0 \\ 0 & 0 & 1 & 0 \\
			0 & 0 & 0 & 1 \\ 1 & 0 & 0 & 0 \end{array} \right)
\]
as studied by Field and Swift \cite{FieldSwift:stationary}.  The
cubic truncation of the general $\bar{G}$-equivariant vector field
is given by
\begin{eqnarray*}
  \dot{x}_1 & = & (\lambda+a_1x_1^2+a_2x_2^2+a_3x_3^2+a_4x_4^2)  x_1 \\
  \dot{x}_2 & = & (\lambda+a_1x_2^2+a_2x_3^2+a_3x_4^2+a_4x_1^2)  x_2 \\
  \dot{x}_3 & = & (\lambda+a_1x_3^2+a_2x_4^2+a_3x_1^2+a_4x_2^2)  x_3 \\
  \dot{x}_4 & = & (\lambda+a_1x_4^2+a_2x_1^2+a_3x_2^2+a_4x_3^2)  x_4
\end{eqnarray*}
First observe that if all the $a_i=-1$, then in the flow of the vector
field, for $\lambda<0$ the origin is globally attracting, and for
$\lambda>0$ there is an invariant attracting sphere of radius
$\sqrt{\lambda}$. The sphere is normally hyperbolic.  If the $a_i$ are
close to $-1$, then there will be an invariant topological $3$-sphere
in the post bifurcation flow (The Invariant Sphere Theorem
\cite{Field:equivariantbreaking}).

We may now ask about the dynamics on the invariant $3$-sphere.  We
rescale the phase space variables by $\sqrt{\lambda}$ and the
independent variable by $1/\lambda$ to blow up the sphere, which is
equivalent to setting $\lambda=1$ in the above equations.  Varying the
parameters $a_i$, we study the flow on the sphere realizing that all
the dynamics on the sphere represent possible post bifurcation
behavior of the vector field.  Structurally stable objects give
persistent post bifurcation behavior for $\bar{G}$-equivariant
families.  All indications imply that this system has no chaotic
behavior, but consider $G \subset \bar{G}$, the elements of the group
$G$ which are orientation preserving.  A new term must be added to the
equivariant vector field which results in
\begin{eqnarray*}
  \dot{x}_1 & = & (\lambda+a_1x_1^2+a_2x_2^2+a_3x_3^2+a_4x_4^2)  x_1 +
e x_2 x_3 x_4 \\
  \dot{x}_2 & = & (\lambda+a_1x_2^2+a_2x_3^2+a_3x_4^2+a_4x_1^2)  x_2 -
e x_1 x_3 x_4 \\
  \dot{x}_3 & = & (\lambda+a_1x_3^2+a_2x_4^2+a_3x_1^2+a_4x_2^2)  x_3 +
e x_1 x_2 x_4 \\
  \dot{x}_4 & = & (\lambda+a_1x_4^2+a_2x_1^2+a_3x_2^2+a_4x_3^2)  x_4 -
e x_1 x_2 x_3
\end{eqnarray*}
The geometry associated with this group action appears to be very rich
and was first examined by Guckenheimer and Worfolk
\cite{GuckenheimerWorfolk:instant}.  For $a_i \approx -1$ and $\lambda
>0$, there is an invariant, attracting, topological sphere in the flow
on which we wish to study the dynamics.  Also, as in the earlier
examples, there are structurally stable heteroclinic cycles given by
the intersection of the invariant sphere and the invariant
$x_i-x_{i+1}$ orthogonal coordinate planes.

The geometry associated with this group action appears to be very
rich.  We may think of objects as bifurcating from the heteroclinic
cycles as they are broken by the introduction of fixed points in the
cycles.  A numerical study reveals unmistakable signs of chaotic
behavior: period doubling cascades and
\u{S}ilnikov homoclinic orbits.  This leads to the conclusion
that the flow on the invariant sphere can contain chaotic attractors.
All hyperbolic invariant sets are possible post bifurcation behavior,
hence we can expect to bifurcate directly from a trivial equilibrium
to a chaotic attractor of small amplitude.  We are currently applying
the techniques presented for the analysis of homoclinic bifurcations
to the task of analytically proving the existence of chaotic behavior
in a neighborhood of the heteroclinic cycles.  This would verify the
following conjecture formulated from the numerical exploration.

\begin{conjecture}
There is an open region $\cal{U}$ in the space of one parameter
families of vector fields on $R^4$ equivariant with respect to $G$,
such that $X_{\lambda} \in \cal{U}$ implies that a bifurcation of the
trivial equilibrium point occurs in $X_{\lambda}$ at $\lambda =
\lambda_0$ that produces ``instant chaos''.  This means that if $U$ is
a neighborhood of the origin in $R^4$, then there is an $\epsilon>0$
such that $X_{\lambda}$ has a chaotic hyperbolic invariant set
contained in $U$ for $0<\lambda-\lambda_0<\epsilon$.
\end{conjecture}

\section{Acknowledgements}
The authors would like to thank Allen Back for his help with the
computation of two-dimensional manifolds and Mark Myers for his help
with the section on Hopf bifurcations.

\section{Captions for Figures}

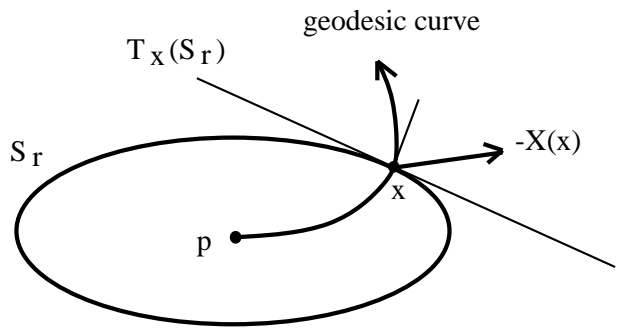
\noindent Figure \ref{geodesic}: Evolve the stable manifold of $p$ using
geodesic curves.
\vskip 0.25in

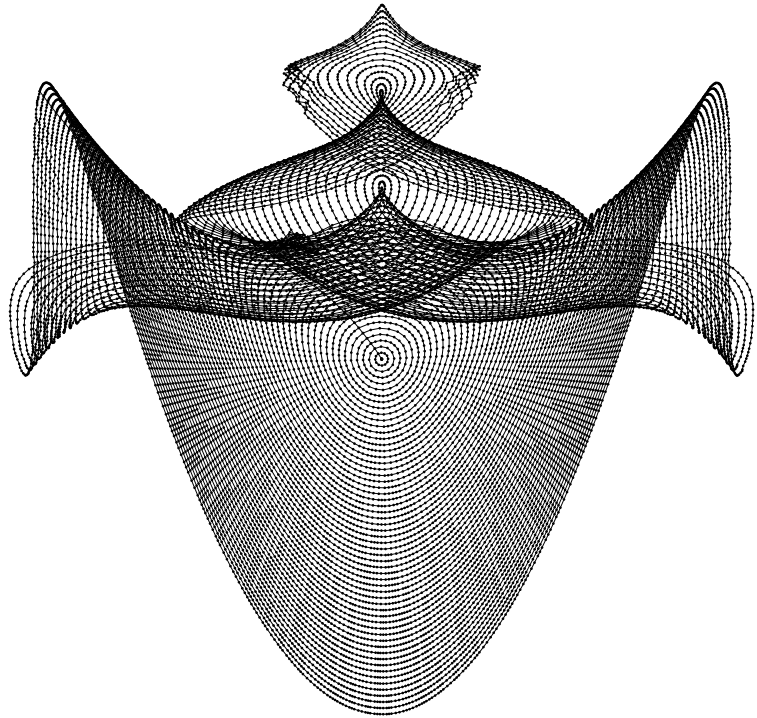
\noindent Figure \ref{twodman}: The stable manifold of the origin
in the Lorenz system projected onto the x-z plane.
\vskip 0.25in

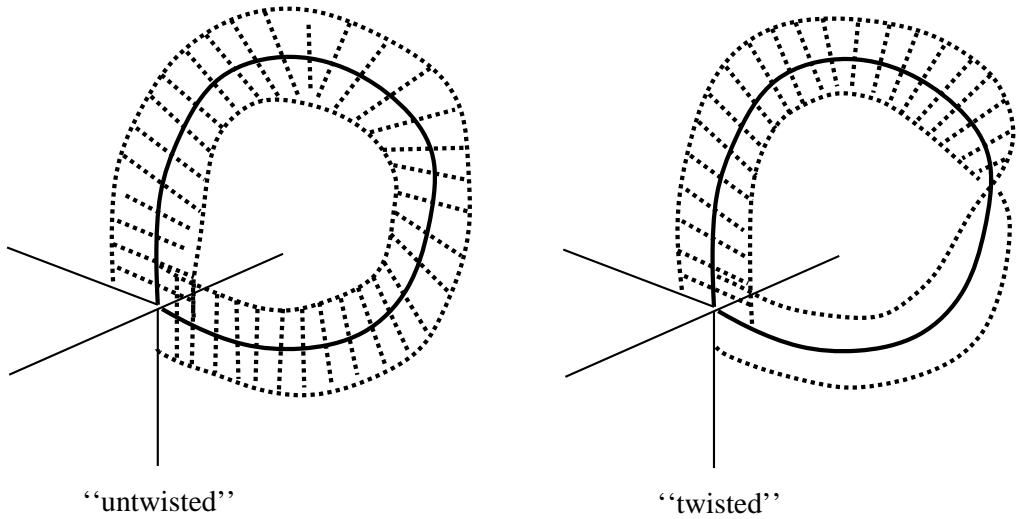
\noindent Figure \ref{homo}: Depiction of section of $W^{s+}\cap W^{u+}$
for typical homoclinic orbits.
\vskip 0.25in

\noindent Figure \ref{focal}: Unifocal and bifocal homoclinic orbits.
\vskip 0.25in

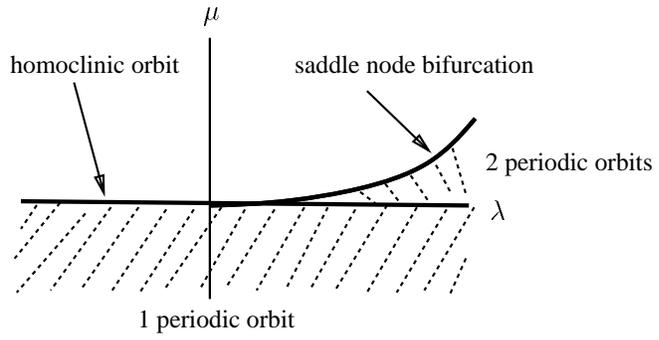
\noindent Figure \ref{res}: Bifurcation diagram for an untwisted resonant
homoclinic orbit.
\vskip 0.25in

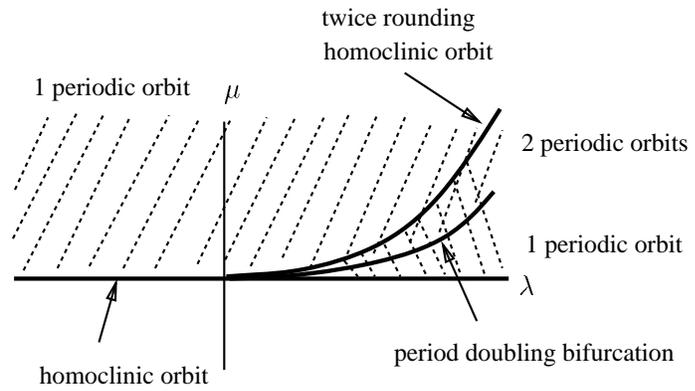
\noindent Figure \ref{restwist}: Bifurcation diagrams for twisted resonant
homoclinic orbit.
\vskip 0.25in

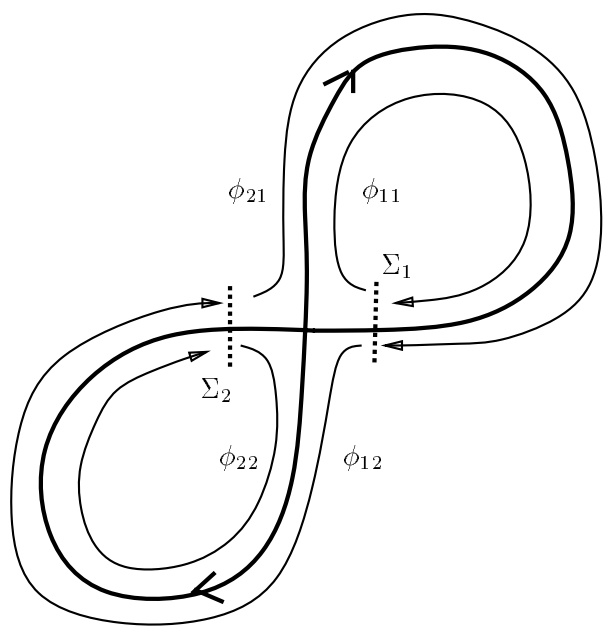
\noindent Figure \ref{gluing}: Return maps for the planar gluing bifurcation.
\vskip 0.25in

\noindent Figure \ref{hh1}: Bifurcation diagram for the Hodgkin and
Huxley equations.
\vskip 0.25in

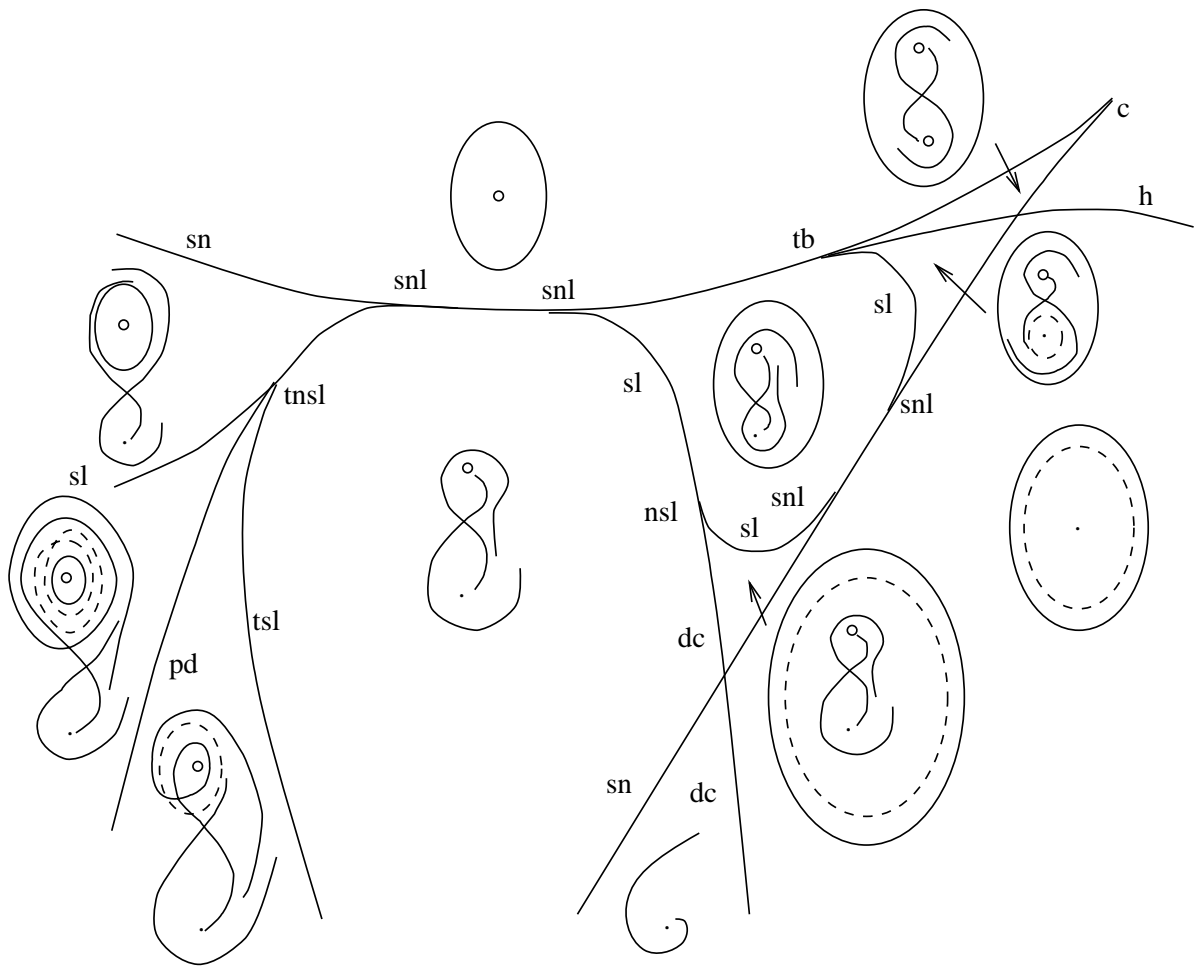
\noindent Figure \ref{hh2}: Bifurcation diagram for the Hodgkin and
Huxley equations, with $\bar{g}_K=12\; {\rm mS}/{\rm cm}^2$.


\vskip 0.25in

\clearpage

\begin{thebibliography}{10}

\bibitem{ArgoulArneodo:experimental}
F.~Argoul, A.~Arneodo, and P.~Richetti.
\newblock Experimental evidence for homoclinic chaos in the
  {B}elousov-{Z}habotinskii reaction.
\newblock {\em Phys. Lett. A}, 120(6):269--275, 1987.

\bibitem{Arnold:geommethods}
V.~I. Arnold.
\newblock {\em Geometrical Methods in the Theory of Ordinary Differential
  Equations}.
\newblock Springer-Verlag, New York, 1988.

\bibitem{dstool:AMS}
A.~Back, J.~Guckenheimer, M.~Myers, F.~Wicklin, and P.~Worfolk.
\newblock dstool: Computer assisted exploration of dynamical systems.
\newblock {\em Notices Amer. Math. Soc.}, 39(4):303--309, 1992.

\bibitem{BaesensGuckenheimer:three}
C.~Baesens, J.~Guckenheimer, S.~Kim, and R.~Mackay.
\newblock Three coupled oscillators: Mode-locking, global bifurcations and
  toroidal chaos.
\newblock {\em Phys. D}, 49:387--475, 1991.

\bibitem{Belyakov:bifurcation}
L.~Belyakov.
\newblock Bifurcation of systems with homoclinic curve of a saddle-focus with
  saddle quantity zero.
\newblock {\em Mat. Zametki}, 36(5):681--689, 1985.

\bibitem{BenedicksCarleson}
M.~Benedicks and L.~Carleson.
\newblock The dynamics of the {H}\'{e}non map.
\newblock {\em Ann. of Math.}, 133:73--169, 1991.

\bibitem{Beyn}
W.~J. Beyn.
\newblock The numerical computation of connecting orbits in dynamical systems.
\newblock {\em IMA J. Numer. Anal.}, 9:169--181, 1990.

\bibitem{BoschSimo:attractors}
M.~Bosch and C.~Simo.
\newblock Attractors in a \u{S}ilnikov-{H}opf scenario and a related
  one-dimensional map.
\newblock {\em Phys. D}, 62:217--229, 1993.

\bibitem{ChowDeng:resonant}
S.-N. Chow, B.~Deng, and B.~Fiedler.
\newblock Homoclinic bifurcation at resonant eigenvalues.
\newblock {\em J. Dynamics Differential Equations}, 2(2):177--244, 1990.

\bibitem{ChowDeng:bifurcation}
S.-N. Chow, B.~Deng, and D.~Terman.
\newblock The bifurcation of homoclinic and periodic orbits from two
  heteroclinic cycles.
\newblock {\em SIAM J. Math. Anal.}, 21(1):179--204, 1990.

\bibitem{CL}
S.~N. Chow and X.~B. Lin.
\newblock Bifurcation of a homoclinic orbit with a saddle-node equilibrium.
\newblock {\em Differential Integral Equations}, 3:435--466, 1990.

\bibitem{Deng:nonhyperbolic}
B.~Deng.
\newblock Homoclinic bifurcations with nonhyperbolic equilibria.
\newblock {\em SIAM J. Math. Anal.}, 21(3):693--720, 1990.

\bibitem{Deng:countable}
B.~Deng.
\newblock The bifurcations of countable connections from a twisted homoclinic
  loop.
\newblock {\em SIAM J. Math. Anal.}, 22(3):653--679, 1991.

\bibitem{Deng:twisting}
B.~Deng.
\newblock Homoclinic twisting bifurcation and cusp horseshoe maps.
\newblock Preprint, Dec. 1991.

\bibitem{Dulac}
M.~H. Dulac.
\newblock Sur les cycles limites.
\newblock {\em Bull. Soc. Math. Anal.}, 51:45--188, 1923.

\bibitem{Dumortier}
F.~Dumortier, R.~Roussarie, and J.~Sotomayor.
\newblock Generic 3-parameter families of vector fields on the plane, unfolding
  a singularity with nilpotent linear part. {T}he cusp case of codimension 3.
\newblock {\em Ergodic Theory Dynamical Systems}, 7:375--413, 1987.

\bibitem{Field:equivariant2}
M.~Field.
\newblock Equivariant dynamics.
\newblock {\em Contemp. Math.}, 56:69--96, 1986.

\bibitem{Field:equivariantbreaking}
M.~Field.
\newblock Equivariant bifurcation theory and symmetry breaking.
\newblock {\em J. Dynamics Differential Equations}, 1(4):369--421, 1989.

\bibitem{FieldSwift:stationary}
M.~Field and J.~Swift.
\newblock Stationary bifurcation to limit cycles and heteroclinic cycles.
\newblock {\em Nonlinearity}, 4:1001--1043, 1991.

\bibitem{FowlerSparrow:bifocal}
A.~C. Fowler and C.~T. Sparrow.
\newblock Bifocal homoclinic orbits in four dimensions.
\newblock {\em Nonlinearity}, 4:1159--1182, 1991.

\bibitem{Friedman}
M.~J. Friedman.
\newblock Numerical analysis and accurate computation of heteroclinic orbits in
  the case of center manifolds.
\newblock To appear in {\em J. Dynamics Differential Equations}, 1992.

\bibitem{DF}
M.~J. Friedman and E.~J. Doedel.
\newblock Computational methods for global analysis of homoclinic and
  heteroclinic orbits: a case study.
\newblock To appear in {\em J. Dynamics Differential Equations}, 1992.

\bibitem{Fuller}
A.~T. Fuller.
\newblock Conditions for a matrix to have only characteristic roots with
  negative real parts.
\newblock {\em J. Math. Anal. Appl.}, 23:71--98, 1968.

\bibitem{Gaspard:generation}
P.~Gaspard.
\newblock Generation of a countable set of homoclinic flows through
  bifurcation.
\newblock {\em Phys. Lett. A}, 97(1):1--4, 1983.

\bibitem{GaspardKapral:homoclinic}
P.~Gaspard, R.~Kapral, and G.~Nicolis.
\newblock Bifurcation phenomena near homoclinic systems: {A} two-parameter
  analysis.
\newblock {\em J. Statist. Phys.}, 35(5/6):697--727, 1984.

\bibitem{GaspardWang:homoclinic}
P.~Gaspard and X.-J. Wang.
\newblock Homoclinic orbits and mixed-mode oscillations in far-from-equilibrium
  systems.
\newblock {\em J. Statist. Phys.}, 48(1):151--199, 1987.

\bibitem{Glendinning}
P.~Glendinning.
\newblock Travelling wave solutions near isolated double-pulse solitary waves
  of nerve axon equations.
\newblock {\em Phys. Lett. A}, 121:411--413, 1987.

\bibitem{Glendinning:subsidiary}
P.~Glendinning.
\newblock Subsidiary bifurcations near bifocal homoclinic orbits.
\newblock {\em Math. Proc. Cambridge Philos. Soc.}, 105:597--605, 1989.

\bibitem{GS:h}
P.~Glendinning and C.~Sparrow.
\newblock Local and global behavior near homoclinic orbits.
\newblock {\em J. Stat. Phys.}, 34(5/6):645--696, 1984.

\bibitem{GlendinningSparrow:tpoints}
P.~Glendinning and C.~Sparrow.
\newblock T-points: A codimension two heteroclinic bifurcation.
\newblock {\em J. Stat. Phys.}, 43(3/4):479--488, 1986.

\bibitem{GlendinningTresser:hyperchaos}
P.~Glendinning and C.~Tresser.
\newblock Heteroclinic loops leading to hyperchaos.
\newblock {\em J. Physique Lett.}, 46(8):347--352, 1985.

\bibitem{GG}
M.~Golubitsky and V.~Guillemin.
\newblock {\em Stable Mappings and Their Singularities}.
\newblock Springer-Verlag, New York, 1973.

\bibitem{GSS}
M.~Golubitsky, I.~Stewart, and D.~Shaeffer.
\newblock {\em Singularities and Groups in Bifurcation Theory}, volume~II.
\newblock Springer-Verlag, New York, 1988.

\bibitem{GPD}
J.~Guckenheimer.
\newblock Multiple bifurcation problems for chemical reactors.
\newblock {\em Phys. D}, 20:1--20, 1986.

\bibitem{GuckenheimerHolmes:heteroclinic}
J.~Guckenheimer and P.~Holmes.
\newblock Structurally stable heteroclinic cycles.
\newblock {\em Math. Proc. Cambridge Philos. Soc.}, 103:189--192, 1988.

\bibitem{GH}
J.~Guckenheimer and P.~J. Holmes.
\newblock {\em Nonlinear Oscillations, Dynamical Systems, and Bifurcations of
  Vector Fields}.
\newblock Springer-Verlag, New York, 1983.

\bibitem{GW}
J.~Guckenheimer and R.~Williams.
\newblock Structural stability of {L}orenz attractors.
\newblock {\em Publ. Math. IHES}, 50:59--72, 1979.

\bibitem{GuckenheimerWorfolk:instant}
J.~Guckenheimer and P.~Worfolk.
\newblock Instant chaos.
\newblock {\em Nonlinearity}, 5:1211--1222, 1992.

\bibitem{Hass2}
B.~D. Hassard and L.-J. Shiau.
\newblock Isolated periodic solutions of the {H}odgkin-{H}uxley equations.
\newblock {\em J. Theoret. Biol.}, 136:267--280, 1989.

\bibitem{Henon}
M.~H\'{e}non.
\newblock A two-dimensional mapping with a strange attractor.
\newblock {\em Comm. Math. Phys.}, 50:69--77, 1976.

\bibitem{HirschbergKnobloch:silnikovhopf}
P.~Hirschberg and E.~Knobloch.
\newblock {S}ilnikov-{H}opf bifurcation.
\newblock {\em Phys. D}, 62:202--216, 1993.

\bibitem{HH52b}
A.~L. Hodgkin and A.~F. Huxley.
\newblock Current carried by sodium and potassium ions through the membrane of
  the giant axon of loligo.
\newblock {\em J. Physiology}, 116:449--472, 1952.

\bibitem{HH52}
A.~L. Hodgkin and A.~F. Huxley.
\newblock A quantitative description of membrane current and its applications
  to conduction and excitation in nerve.
\newblock {\em J. Physiology}, 117:500--544, 1952.

\bibitem{holden}
A.~V. Holden, M.~A. Muhamad, and A.~K. Schierwagen.
\newblock Repolarizing currents and periodic activity in nerve membrane.
\newblock {\em J. Theoret. Neurobiol.}, 4:61--71, 1985.

\bibitem{GGT}
P.~Glendinning J.~M.~Gambaudo and C.~Tresser.
\newblock Collage de cycles et suites de {F}arey.
\newblock {\em C. R. Acad. Sci. Paris}, 299:711--714, 1984.

\bibitem{Keller}
H.~B. Keller.
\newblock Numerical solution of bifurcation and nonlinear eigenvalue problems.
\newblock In P.~Rabinowitz, editor, {\em Applications of Bifurcations Theory},
  pages 359--384. Academic Press, 1977.

\bibitem{Kokubu:homoclinic}
H.~Kokubu.
\newblock Homoclinic and heteroclinic bifurcations of vector fields.
\newblock {\em Japan J. Appl. Math}, 5:455--501, 1988.

\bibitem{Lab}
I.~S. Labouriau.
\newblock Degenerate {H}opf bifurcation and nerve impulse, {P}art {II}.
\newblock {\em SIAM J. Math. Anal.}, 20:1--12, 1989.

\bibitem{Loos}
R.~Loos.
\newblock Generalized polynomial remainder sequences.
\newblock In G.E~Collins B.~Buchberger and R.~Loos, editors, {\em Computer
  Algebra - Symbolic and Algebraic Computation}, pages 115--137.
  Springer-Verlag, 1982.

\bibitem{Lorenz}
E.~N. Lorenz.
\newblock Deterministic non-periodic flow.
\newblock {\em J. Atmospheric Sci.}, 20:130--141, 1963.

\bibitem{Lukyanov:bifurcations}
V.~Luk'yanov.
\newblock Bifurcations of dynamical systems with a saddle-point-separatrix
  loop.
\newblock {\em J. Differential Equations}, 18:1049--1059, 1983.

\bibitem{MV}
L.~Mora and M.~Viana.
\newblock Abundance of strange attractors.
\newblock Preprint, 1991.

\bibitem{PdM}
J.~Palis and W.~de~Melo.
\newblock {\em Geometric Theory of Dynamical Systems}.
\newblock Springer-Verlag, New York, 1982.

\bibitem{RichettiArgoul:intermittency}
R.~Richetti, F.~Argoul, and A.~Arneodo.
\newblock Type-{II} intermittency in a periodically driven nonlinear
  oscillator.
\newblock {\em Phys. Rev. A}, 34(1):726--729, 1986.

\bibitem{R&M}
J.~Rinzel and R.~N. Miller.
\newblock Numerical solutions of the {H}odgkin-{H}uxley equations.
\newblock {\em Math. Biosc.}, 49:27--59, 1980.

\bibitem{Schecter:saddlenode}
S.~Schecter.
\newblock The saddle-node separatrix-loop bifurcation.
\newblock {\em SIAM J. Math. Anal.}, 18(4):1142--1156, 1987.

\bibitem{Schecter:simultaneous}
S.~Schecter.
\newblock Simultaneous equilibrium and heteroclinic bifurcation of planar
  vector fields via the {M}elnikov integral.
\newblock {\em Nonlinearity}, 3:79--99, 1990.

\bibitem{Schecter:singularity}
S.~Schecter.
\newblock {$C^1$} singularity theory and heteroclinic bifurcation with a
  distinguished parameter.
\newblock {\em J. Differential Equations}, 99:306--341, 1992.

\bibitem{Schecter2}
S.~Schecter.
\newblock Numerical computation of saddle-node homoclinic bifurcation points.
\newblock to appear in SIAM J. Numer. Anal., 1993.

\bibitem{Silnikov:countable}
L.~P. Silnikov.
\newblock A case of the existence of a countable number of periodic motions.
\newblock {\em Soviet Math. Dokl.}, 6:163--166, 1965.

\bibitem{Silnikov:existence}
L.~P. Silnikov.
\newblock The existence of a denumerable set of periodic motions in
  four-dimensional space in an extended neighborhood of a saddle-focus.
\newblock {\em Soviet Math. Dokl.}, 8(1):54--58, 1967.

\bibitem{Silnikov:generation}
L.~P. Silnikov.
\newblock On the generation of a periodic motion from trajectories doubly
  asymptotic to an equilibrium state of saddle type.
\newblock {\em Math. USSR-Sb.}, 6(3):427--438, 1968.

\bibitem{Silnikov:contribution}
L.~P. Silnikov.
\newblock A contribution to the problem of the structure of an extended
  neighborhood of a rough equilibrium of saddle-focus type.
\newblock {\em Math. USSR-Sb.}, 10(1):91--102, 1970.

\bibitem{Stanley:invariants}
R.P. Stanley.
\newblock Invariants of finite groups and their applications to combinatorics.
\newblock {\em Bull. Amer. Math. Soc.}, 1(3):475--511, 1979.

\bibitem{Terman:transition}
D.~Terman.
\newblock The transition from bursting to continuous spiking in excitable
  membrane models.
\newblock {\em J. Nonlinear Science}, 2(2):135--182, 1992.

\bibitem{Tresser:silnikov}
C.~Tresser.
\newblock About some theorems by {L.P. \u{S}ilnikov}.
\newblock {\em Ann. Inst. H. Poincar\'{e} Sect. A}, 40(4):441--461, 1984.

\end{thebibliography}
\end{document}